\documentclass[useAMS,usenatbib,referee]{mnras}
\usepackage{amsfonts}
\usepackage{amsmath}
\usepackage{amssymb}

\usepackage[dvips]{graphicx}
\usepackage{color}
\usepackage{multirow}
\def\bea{\begin{eqnarray}}
\def\ena{\end{eqnarray}}
\newcommand{\mr}[1]{\mathrm{#1}}

\title[Search for isolated stellar mass black holes]
{Search for isolated Galactic Centre stellar mass black holes in the IR and sub-mm range}

\author[P. B. Ivanov,  V. N. Lukash, S. V. Pilipenko, M. S. Pshirkov]
{P.B. Ivanov$^{1}$\thanks{E-mail: pbi20@cam.ac.uk (PBI)},
 V. N. Lukash $^{1}$\thanks{E-mail: lukash@asc.rssi.ru (VNL)},
S. V. Pilipenko $^{1}$\thanks{E-mail: sergey.f77@gmail.com  (SVP)}
and  M. S. Pshirkov$^{2}$\thanks{E-mail:pshirkov@sai.msu.ru (MSP, corresponding author)}\\
$^{1}$Astro Space Centre, P.N. Lebedev Physical Institute, 4/32
Profsoyuznaya Street, Moscow, 117810, Russia\\$^{2}$
Sternberg Astronomical Institute, Moscow State University, 119992, 
Universitetskiy Prospekt, 13, Moscow, Russia
}

\begin{document}

\date{}

\pubyear{2019}

\maketitle

\label{firstpage}

\begin{abstract}
We investigate a possibility to find an accreting isolated black hole (IBH) with 
mass $1-100\mr{M}_{\odot}$ within Central 
Galactic Molecular Zone (CMZ) in the submillimetre and IR spectral range with help of planned
space observatories James Webb Space Telescope (JWST) and Millimetron (MM). We assume 
the spherical mode of accretion. We develop 
the simplest possible quantitative  model of the formation of radiation spectrum in this range due to synchrotron emission and show that it fully agrees with the more complicated model of Ipser and 
Price 1982 for expected values of accretion rate.

If a substantial fraction of LIGO events was caused by merger of primordial black holes, the JWST would be able to find them  provided that there is a cusp in distribution of dark matter in the Galaxy and that  the accretion efficiency parameter $\lambda $ defined as the ratio of accretion rate onto IBH
to its Bondi-Hoyle-Lyttleton value is larger than  $\sim 10^{-2}$. A comparable amount of IBHs is also predicted
by recent models of their formation  due to stellar evolution. MM capabilities are 
hampered by the effect of confusion due to distant submillimetre galaxies, but it can also
be used for such purposes if the confusion effect is properly dealt with. In case of efficient accretion with $\lambda \sim 1$, both instruments could detect IBHs even when their number densities are as low as $10^{-6}~\mr{pc}^{-3}$. 
 
\end{abstract}


\section{Introduction}
\label{sec:intro}

Isolated stellar mass black holes (IBHs) can be either formed as a result of stellar evolution (see e.g. \citep{Shapiro1983}, 
\citep{Eldridge2011}  and references therein) or be primordial (e.g.  \citep{Carr2016} and reference therein). Recently the interest in finding such black holes has grown
significantly since the first detection of gravitational wave signal  GW150914  originated from a merger of two $\sim30~\mr{M}_{\odot}$
black holes followed by a number of several similar events. There are some arguments suggesting their primordial origin. For instance,  \citep{Sasaki2016} estimate dark matter fraction $f$ \footnote{$f=\rho_{PBH}/\rho_{DM}$, where $\rho_{PBH}$ and $\rho_{DM}$ are mass densities of primordial black holes and dark matter, respectively.}
sufficient to explain the observed merger rate to be of the order of $10^{-3}$.

Unlike black holes in binary systems, IBH are expected to be quite dim. So far, several authors have investigated a possibility 
of their search through microlensing (e.g. \citep{Bennett2002}), or observations in radio (e.g. \citep{Gag}, \cite{Maccarone}), IR (e.g. \citep{Mc1985}, \citep{Campana1993}), optical (e.g. \citep{Chisholm2003}) bands, X-rays (e.g. \citep{Fujita1998},  \citep{Agol2002}, \citep{Fender}, \citep{Gag}, \cite{Hektor}) and gamma rays (e.g. \citep{Armitage1999}), see also references in those papers, with no confirmed candidates having yet been found.

Up to now only IBHs in the closest neighbourhood of the Sun or in the nearest molecular clouds have 
been considered as detectable sources, e.g. \cite{Fender}. One of possible reason for the non-detection of IBHs  would be small spatial volumes of these sites, which impedes finding objects with a low abundance. It is clear that much larger test volumes 
and new observational facilities with better sensitivities are needed for future attempts to observe IBHs.

A natural candidate for such a site would be a region close to the Galactic Centre, say, the central region with diameter $100-200$~pc.
Firstly, the interstellar medium is sufficiently dense there, with the number density of hydrogen of the order of $10^2~\mr{cm}^{-3}$ much larger than the mean number density in our Galaxy $\sim 1~\mr{cm}^{-3}$, 
thus, enhancing the accretion
rate onto an IBH. Secondly, both recent calculations of spatial distribution of IBHs born as a result of stellar 
evolution \citep{Tsuna2018} and the assumption that they are of primordial origin and should, accordingly, trace the distribution of dark matter suggest that they should concentrate near the Galactic Centre (GC), thus, pointing to this site as a natural place to
look for IBHs.  Thirdly, the small angular size of this region implies that it can be easily surveyed during a reasonable observational time.  

The search for IBHs in GC is obstructed by its large distance from Earth and large expected average velocities of IBHs with respect to the GC interstellar medium\footnote{The latter is due to the fact that accretion rate onto an IBH should be inversely proportional to the third power of $v$, see eq. (\ref{BH}).}, $v \sim 200~\mr{km~s^{-1}}$. 
However, the next decade will see the launches of two space observatories 
with unprecedented sensitivities in the IR and submillimetre range, James Webb Space Telescope (JWST, see e.g. https://www.jwst.nasa.gov/) 
and Millimetron (MM, see e.g. http://millimetron.ru/index.php/en/, \citep{Kardashev2014} 
and \citep{Ivanov}). Both instruments will reach the sensitivity of the order
of $10^{-5}$~Jy, making it possible to detect extremely faint objects and, thus, suggesting them for the search of IBHs in GC.

It is the purpose of this paper to explore a possibility of detection of IBHs in GC with help of these observatories. Since the physics
of accretion and formation of radiation spectrum in this waveband are rather poorly known, we confine ourselves to the simplest quantitative 
model of accretion and production of radiation in an accreting IBH. Namely, we assume that accretion proceeds through  the Bondi-Hoyle-Lyttleton (BHL) mechanism at a rate
\begin{equation}
 \label{BH}
 \dot{M}=4\pi\lambda\frac{G^2M^2\rho}{(v^2+c_s^2)^{3/2}},
\end{equation}
where $G$ is the gravitational constant, $M$ is the mass of a black hole  travelling at  velocity  $v$ through interstellar medium of density $\rho$ and sound speed $c_s$, $\lambda $ is an efficiency factor, which takes into account that 
the accretion rate could be smaller than its 'canonical' BHL value, {see e.g. \cite{Perna}, \cite{Fender} and references therein.} 
    
We consider only the synchrotron mechanism of radiation emission in the accreting flow, because it is the only mechanism responsible for
production of photons with frequencies in the IR and submillimetre band for expected values of electron number density, temperature
and magnetic field. That an IBH must emit radiation due to this mechanism was first pointed out in the pioneering work \citep{Shvartsman1971}. The first detailed theories of  this process were  proposed in \citep{Shapiro1973}, \citep{BK}, although the former paper considered only
the optically thin regime and neglected possible dissipation of magnetic field, while the latter dealt with a similar case of disc-like structures and  also neglected the effect of dissipation. A detailed numerical calculation of emerging spectrum was done by \citep{IP1982},
who considered gas heating due to dissipation of magnetic field, non-zero opacity due to synchrotron self-absorption and relativistic 
effects\footnote{See also \citep{IP1982} for a discussion of some other early works on the subject.}. In these papers it was assumed
that only thermal electrons contribute to radiation emission. Later, in  \citep{Beskin2005} a possible contribution of non-thermal
electrons was discussed, which changes the spectrum at frequencies higher than those available for observations with MM and JWST, and, therefore, is not important for our purposes as well
as the process of Comptonization of synchrotron radiation.  

We adopt an approach to calculation of the emerging spectrum, which closely follows \citep{IP1982}, making, however, a set of additional simplifications, which reduce the calculations to a small number of simple quadratures. Namely, we assume that the magnetic field energy
density is always proportional to gas gravitational energy density and that all quantities of interest are simple power law functions
of distance from the black hole. Additionally, our model is fully Newtonian. There are two free parameters, the ratio of electron and proton temperatures and inner cut-off radius, which is chosen to be equal to the radius of the circular photon orbit for a non-rotating black hole.
The ratio of temperatures is fixed by the requirement that our calculations give the result for the spectrum as close to the one obtained
by  \citep{IP1982} for particular values of black hole mass and accretion rate. We check that our model, remarkably, is able to reproduce
quantitatively the results of  \citep{IP1982} for a very wide range of black hole masses and sufficiently small accretion rates. This
suggests that, when the accretion rate is small enough, the formation of spectrum does not depend on many details of more complicated 
models. 

We use equation (\ref{BH}) to relate the accretion rate to parameters of Central Molecular Zone (CMZ) of GC assuming  the gas number density  unperturbed by IBHs to be equal to $10^{2}~\mr{cm}^{-3}$, see \citep{Fer} and velocity dispersion $\sigma=200~\mr{km~s^{-1}}$ to be order of velocity dispersion of 
the bulge stars, see \citep{Valenti2018}. The sound speed of the medium is expected to be much smaller than $v$, and, therefore, is set to zero
in (\ref{BH}).  We treat the efficiency parameter $\lambda$ as a free one, and calculate the spectra for a range of
$\lambda $ and black hole masses within the range $1\mr{M}_{\odot} \le M \le 100\mr{M}_{\odot}$ comparing them with the appropriate sensitivity
curves for MM and JWST, taking into account a possible confusion between IBHs and background sources, like sub-mm IR galaxies. Note that all considered accretion rates are inside the region of applicability of our simplified model, see equation (\ref{e10}) below.   

It is shown that MM and JWST are able to detect an IBH within CMZ travelling at a typical speed $\sigma \sim 200~\mr{km~s^{-1}}$ when its mass is larger than $M_{crit}\approx 20~\mr{M}_{\odot}$ and $M_{crit}\approx 30~\mr{M}_{\odot}$, for 
MM and JWST, respectively and $\lambda \sim 1$.   When $\lambda $ gets smaller $M_{crit}$ gets larger, and, for the largest considered $M_{crit}$, the minimal possible value of $\lambda \sim 0.1$ for both instruments.

The detection capabilities of MM are, however, hampered by the confusion limit, which results in
$M_{crit}\approx 90~\mr{M}_{\odot}$ for $\sigma \approx 200~\mr{km~s^{-1}}$. Therefore, one must find a way to disentangle the contributions from distant galaxies and IBHs when using this instrument. This could, perhaps, be done with help of observations at different frequencies and using different variability patterns of IBHs and background sources. It is worth noting that JWST operates at higher frequencies than MM does and, therefore, possesses a much better spatial resolution and is, hence, free from limitations imposed  by the confusion effect.  Taking into account that surveys of CMZ by these instruments are expected in any case we conclude that they will be able to find accreting IBHs when their number density is larger than the inverse volume of CMZ, $\sim 10^{-6}~\mr{pc}^{-3}$, provided that our spectral model is valid and that $\lambda$ is
large enough, $\lambda \sim 0.1-1$.

 On the other hand, when the number of IBHs within CMZ, $N_{IBH}$, is large, there could be black holes travelling with respect to interstellar medium with velocities smaller than $\sigma$. Assuming the Maxwell statistics of velocity distribution we find that in this case it could be possible to detect
IBHs with the accretion efficiency $\lambda \sim 6/N_{IBH}$ depending on their typical mass as well as IBHs with masses down to $1\mr{M}_{\odot}$. 

Our paper is organised as follows. In the next Section we introduce our simplified model of the emission of synchrotron radiation and compare the obtained spectra with those of  \citep{IP1982}. In Section \ref{detection} we find limitations on black hole masses $M$ and efficiency parameter $\lambda$ from the requirement of detection of an IBH in CMZ by MM and JWST. We would like to stress that even in case of non-detection of IBHs by these instruments it will be possible to obtain limitations on their abundance and/or on the value of $\lambda$. We adopt the cgs system of units throughout the text. Note that Sections \ref{spectrum} 
and \ref{detection} are self-contained, so readers who are interested only in astrophysical applications may start reading the text from Section \ref{detection}.   We discuss our results and related issues in Section \ref{discussion}.

\section{The spectral model}
\label{spectrum}

In general, a calculation of the spectrum emerging from an isolated accreting black hole requires an accretion model, which necessarily employs a large number of numerical procedures.   
In what follows we would like to try to eliminate as much numerical work  as possible and develop the simplest model, which is still able to reproduce quantitatively the numerical results reported in \citep{IP1982}. It turns out that such a model can be constructed when accretion rate is sufficiently small. It employs merely a few straightforward numerical quadratures and not only can easily be used by other researchers but also clarifies the most important physical mechanisms leading to the formation of the emission spectrum.  

As we have mentioned, we consider only spherical accretion in this work\footnote{Note that it is straightforward to generalise 
our approach to the case of radiatively inefficient accretion flows.}. Since we are interested only in sub-mm and IR parts of the spectrum,
we consider the synchrotron emission, which is the only radiation mechanism responsible for its formation for gas densities, its 
temperature and values of magnetic field relevant for our systems. We closely follow basic assumptions and notations of \citep{IP1982}
but make a number of additional simplifying assumptions, notably, we consider only a Newtonian problem and treat relativistic effects only in some 'effective' sense, see below.

It is assumed that radial velocity of accreting gas, $v_r$, is proportional to the free fall velocity, $v_r=-{\beta_V}\sqrt{{2GM\over R}}$, where $M$ is the black hole mass,  $R$ is the radial distance and $\beta_{V}$ is a constant. Gas number density, $n$, can be found
from the mass conservation law, ${\dot M}=-4\pi nv_r R^{2}$.  Following  \citep{IP1982} we assume that the energy density of magnetic field, $B$, is a fraction of gravitational energy of infalling gas according to relation 
\begin{equation}
{B^{2}\over 8\pi}=\beta_{B}{GMm_pn\over R},
\label{eq_B}
\end{equation}
where $m_p$ is the proton mass and $\beta_{B}$ is another constant factor.

 Similar to  \citep{IP1982} we assume that only thermal electrons contribute to the formation of the spectrum.
Thus, the emitting electons are distributed over velocities according to the relativistic Maxwellian (or, 
Maxwell-Juttner) distribution. 
The intensity of synchrotron emission is very sensitive to
electron temperature, $T_c$. Unlike  \citep{IP1982} and \citep{Beskin2005} we do not solve any equations describing 
a dependency of $T_{e}$ on $R$, postulating that it is a fraction of a 'natural' proton temperature 
$T_{pr}={m_p c^2R_g \over k_{B}R}$, 
where $k_B$ is the Boltzmann constant and $R_{g}={2GM\over c^2}$ is the gravitational radius. Accordingly, we have 
\begin{equation} T_{e}=\epsilon 10^{12}{1\over r}K,
\label{e1}
\end{equation} where $r=R/R_g$ and $\epsilon \le 1$ is considered to be a constant.
Using the known properties of
the relativistic Maxwellian distribution and equation (\ref{e1}) we estimate the average Lorentz factor
of emitting electrons, $\bar \gamma$, as $\bar \gamma \approx {3k_BT_e\over m_ec^2}\approx 5.1\cdot 10^2{\epsilon \over r}$.

Using these relations it is easy to find the explicit dependencies of $n$ and $B$ on $r$. We have 
\begin{equation} 
n=n_*\dot m m^{-1}\beta_{V}^{-1}r^{-3/2}, \quad n_*\approx 2.5\cdot 10^{10}~\mr{cm}^{-3}, 
\label{e2}
\end{equation}
and 
\begin{equation} 
B=B_*(\beta_{B}\beta_{V}^{-1}\dot m m^{-1})^{1/2}r^{-5/4},
\quad B_*\approx 2.1\cdot 10^{4}~\mr{G},
\label{e3}
\end{equation}    
where  $m={M\over 10\mr{M}_{\odot}}$, $\dot m={\dot M\over 10^{-8}\dot M_{E}}$,
and $\dot M_{E}$ is in turn the Eddington accretion rate calculated for the standard ten per cent radiation efficiency, $\dot M_E\approx 1.4 m\cdot 10^{19}~\mr{g~s^{-1}}$. It is worth noting that  \citep{IP1982} use a quantity $\bar g$ related to  $\dot m$ through $\bar g={1\over 16}\beta_V^{-1}\dot m m^{-1}$.

Luminosity per frequency unit, $L_{\nu}$, due to synchrotron emission reads
\begin{equation} 
L_{\nu}=8\pi^2\int_{R_{min}}^{\infty}dR^{'}{R^{'}}^{2}e^{-\tau(R^{'})}j_{\nu}(R^{'}), \quad \tau (R)=\int^{\infty}_{R} dR^{'} {j_{\nu}\over B_{\nu}},
\label{e4}
\end{equation}   
where $\nu$ is a given frequency, $j_{\nu}$ is the synchrotron emissivity,  $\tau$ is the optical depth, and $B_{\nu}={2\nu^{2}k_{B}T\over c^2}$ is the Planck intensity in the Rayleigh-Jeans limit.

Expression (\ref{e4}) is a simplified version of that of \citep{IP1982}. The latter contains an additional integration 
over the angle $\mu$ between the direction to black hole and the direction of photon propagation. Here we imply that the integrand
does not depend on this variable and that the integration is performed over the angles larger than $\pi/2$ assuming that 
photons with smaller angles are captured by the black hole. This happens close to the radius of the photon sphere around a non-rotating 
black hole $R_{ph}=1.5~R_{g}$, where the bulk of outgoing radiation is formed. Additionally, the expression for the optical depth
is taken over the radial direction contrary to  \citep{IP1982}, where it is evaluated along a particular photon orbit, and 
we neglect all relativistic corrections, effectively taking into account some of them by setting $R_{min}=R_{ph}$, because radiation from smaller radii is strongly redshifted and must be emitted within a progressively smaller range of angles $\mu$ to escape to infinity as $R$ gets smaller than $R_{ph}$. 

Using (\citep{IP1982}) and equations (\ref{e1})-(\ref{e3}) it is easy to obtain explicit expressions for all quantities entering (\ref{e4}). We have
\begin{equation} 
j_{\nu}={3^{1/2}e^3nB\over 8\pi m_e c^2}\Phi(x)=j_*\beta_{B}^{1/2}(\beta_V^{-1}\dot m m^{-})^{1/2}\Phi(x), \quad x=\nu/\nu_{c},
\label{e5}
\end{equation}   
where $e$ and $m_e$ are the electron charge and mass, respectively, 
$j_*=4.9\cdot 10^{-9}~\mr{erg~cm^{-3}~s^{-1}}$ and $\nu_{c}$ is the so-called critical frequency 
\begin{equation} 
\nu_{c}= {3e B (k_{B}T)^2\over 4\pi m_e^3 c^5}=\nu_*\epsilon^2 (\beta_B\beta_V^{-1}\dot m m^{-1})^{1/2}r^{-13/4}, \nu_{*}=2.5\cdot 10^{15}~\mr{Hz},
\label{e6}
\end{equation} 
where we set a characteristic value of sine of 
the pitch angle to be equal to unity. $\Phi(x)$ is given 
by the expression
\begin{equation} 
\Phi(x)=\int^{\infty}_0 dz z^{2}e^{-z}F(x/z^2), \quad F(x)=x\int^{\infty}_{x}dz K_{5/3}(z),
\label{e7}
\end{equation} 
where $K_{5/3}(z)$ is a modified Bessel function of the second kind. 
It is related to the function $\Pi(x)$ introduced in \citep{Pac}
as $\Phi(x)=x\Pi(x)$. The optical depth can be represented as 
\begin{equation} 
\tau=\tau^{*}(\beta_{B}^{-1}\beta_{V}^{-1}\dot m m)^{1/2}{\tilde \nu}^{-2}\int^{\infty}_{r}dr r^{-7/4}\Phi(x), \tau_{*}=7.7\cdot 10^{-9},
\label{e8}
\end{equation}  
where $\tilde \nu =\nu/\nu_{c}(r=1)$. 
 
Substituting (\ref{e5})-(\ref{e8}) in (\ref{e4}) we obtain
\begin{equation} 
L_{\nu}=L_*\beta_{B}^{1/2}(\beta_V \dot m m)^{3/2}\int_{3/2}^{\infty}dr r^{-3/4}e^{-\tau}\Phi(x), \quad L_{*}=10^{13}~\mr{erg~s^{-1}~Hz^{-1}}.
\label{e9}
\end{equation} 

We use expression (\ref{e9}) to compare the results obtained with the help of our simplified model with those
of \citep{IP1982} setting $\beta_{V}=\beta_{B}=1$ as in this paper. After
that, the Ipser and Price model depends only on two parameters, which are
the black hole mass and their accretion factor $\bar g$ proportional to 
our dimensionless accretion rate $\dot m$. The only parameter, which is varied in our model to fit it to that of Ipser 
and Price is the electron temperature factor $\epsilon$, since the temperature is expected to be smaller than the virial one
close to $R_{min}$. We set $\epsilon=0.2$ to obtain a good agreement between the results corresponding to $m=1$ and $\bar g=10^{-4}$
and use other values of $m$ and $\bar g$ to validate our model.
\begin{figure}
\begin{center}
\includegraphics[scale=0.5]{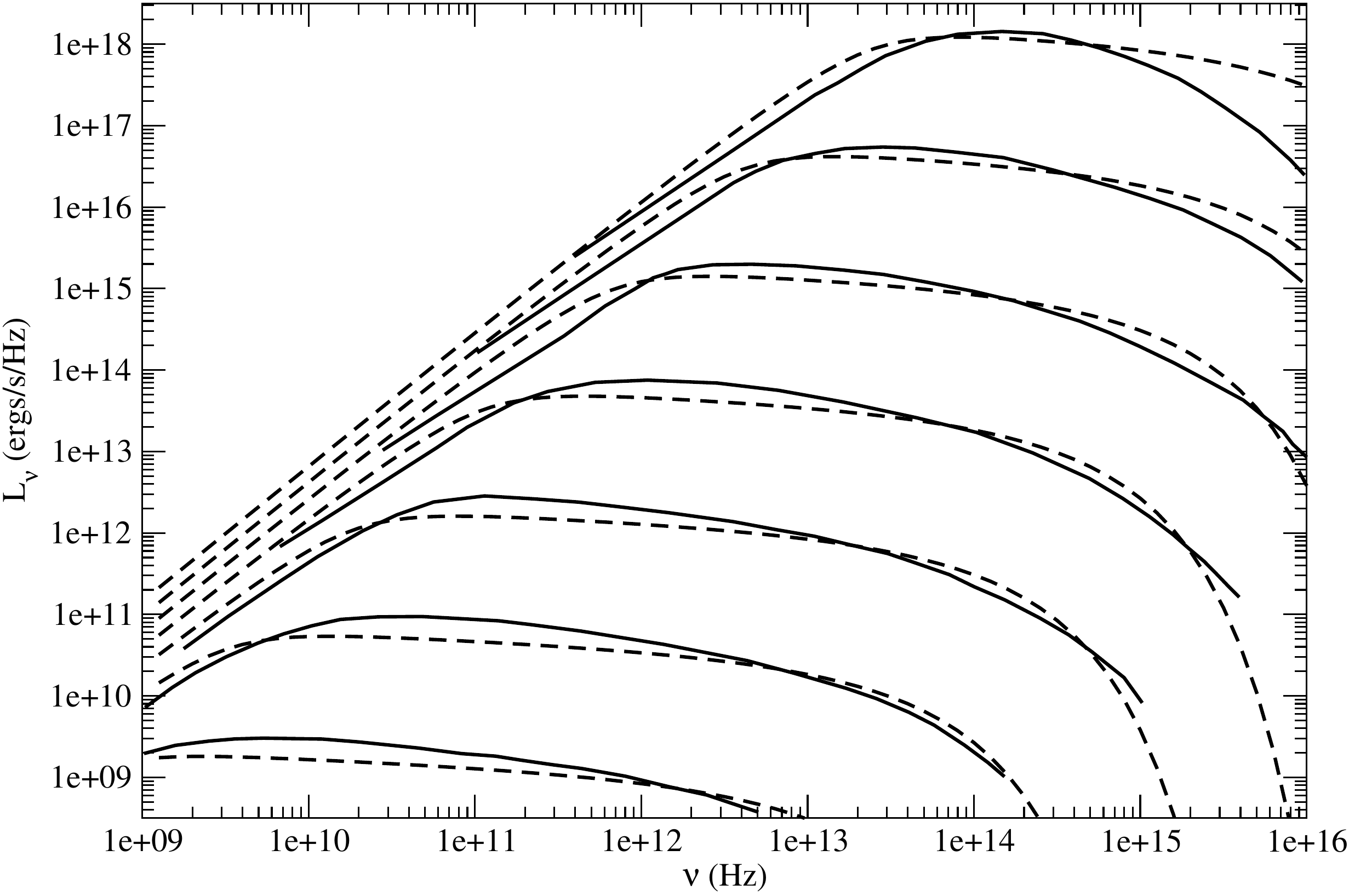}

\end{center}
\caption{Solid lines reproduce the results of Fig. 5a of \citep{IP1982}, which shows $L_{\nu}$ as a function of $\nu$ calculated in their model for $M=10\mr{M}_{\odot}$
and different values of $\bar g$. The corresponding results of our calculations 
are additionally shown as dashed lines.}
\label{Fig1}
\end{figure}  

\begin{figure}
\begin{center}
\includegraphics[scale=0.5]{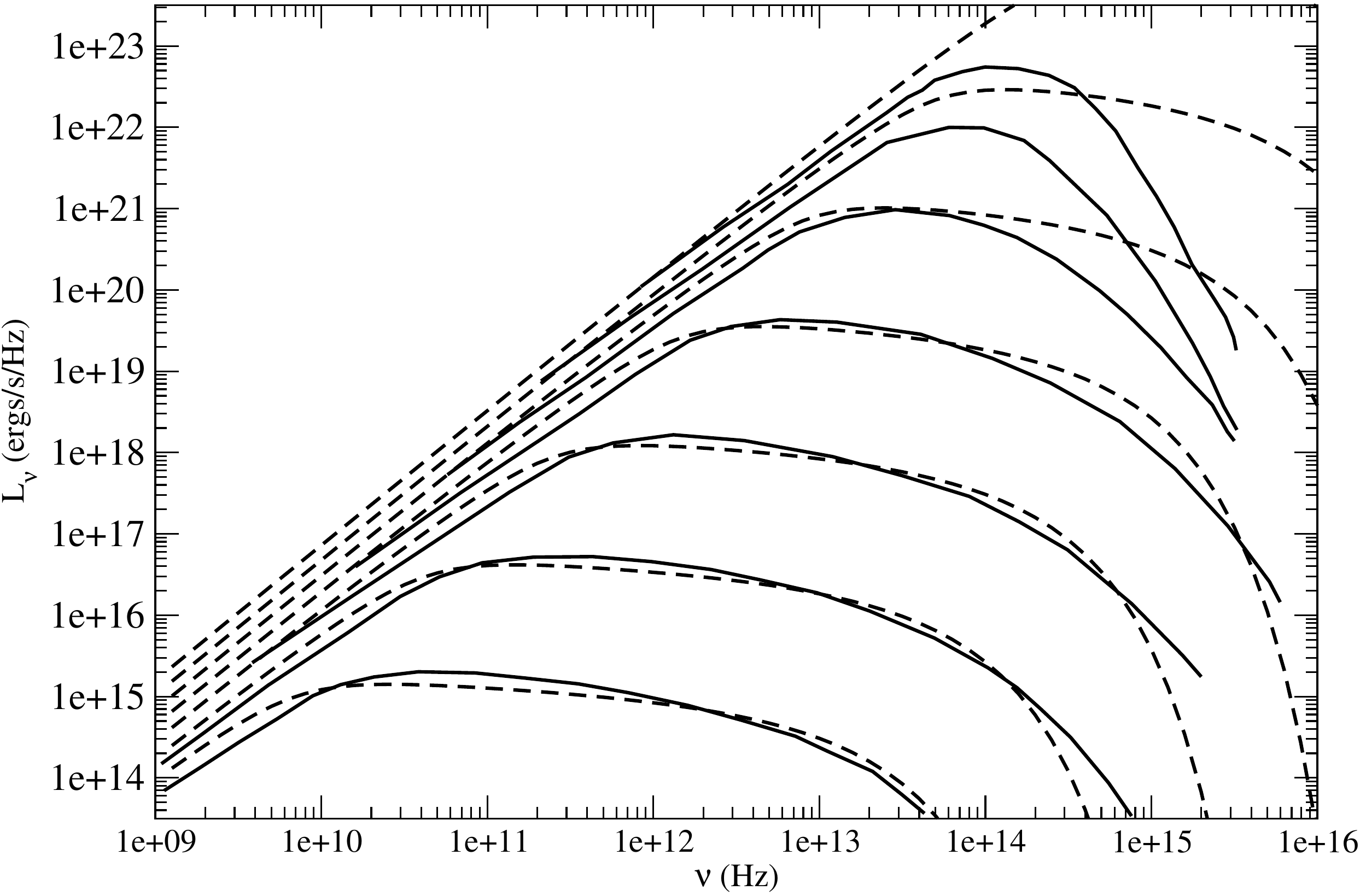}
\end{center}
\caption{Same as \ref{Fig1}, but for $M=10^{3}\mr{M}_{\odot}$, see Fig 5b of (\citep{IP1982}).}
\label{Fig2}
\end{figure} 

\begin{figure}
\begin{center}
\includegraphics[scale=0.5]{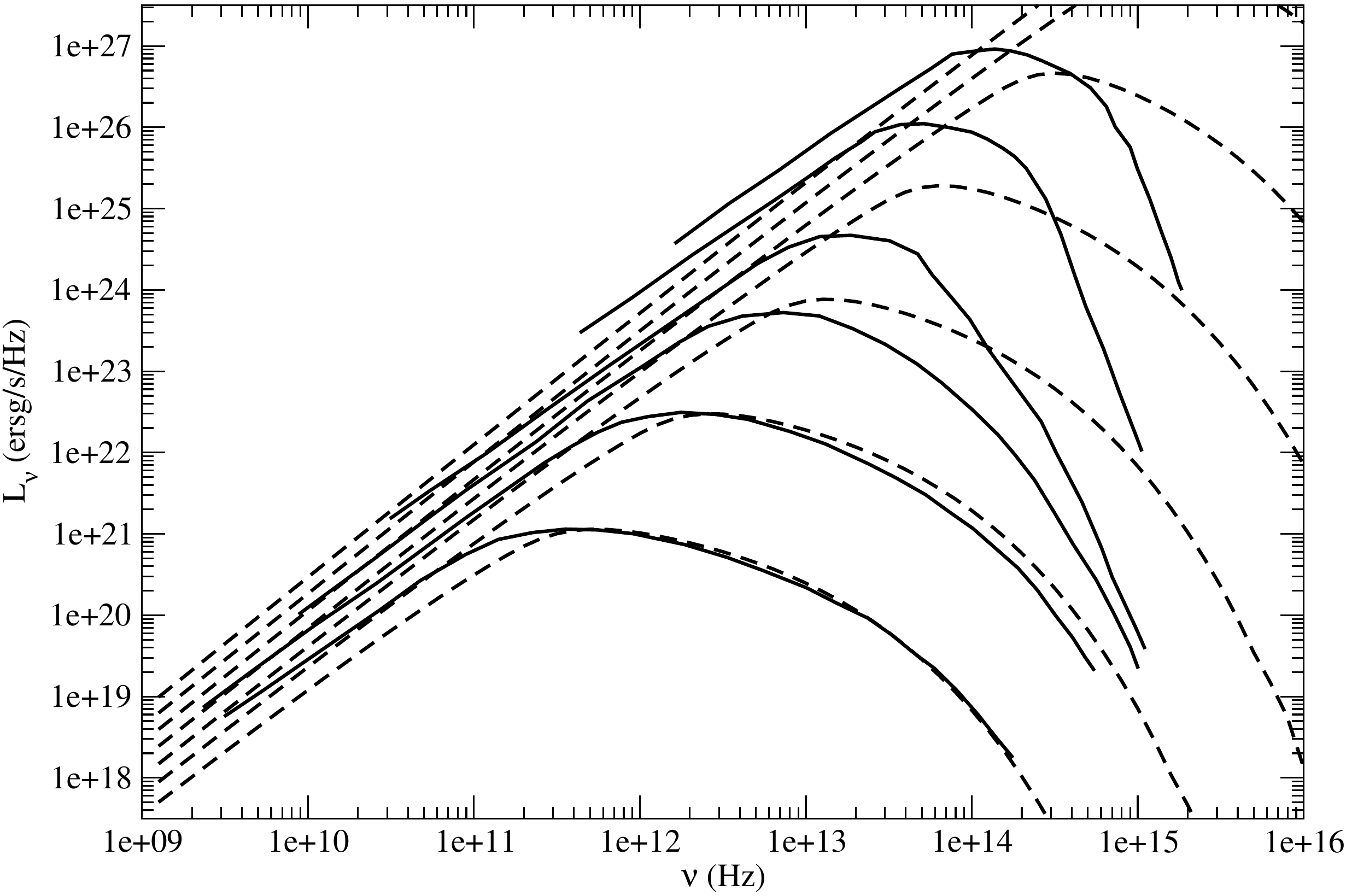}
\end{center}
\caption{Same as \ref{Fig1}, but for $M=10^{5}\mr{M}_{\odot}$, see Fig 5c of (\citep{IP1982}).}
\label{Fig3}
\end{figure} 

The results of comparison are shown in Figs \ref{Fig1}-\ref{Fig3}. As seen from Figs. \ref{Fig1}-\ref{Fig3}, there is a very good agreement between our semi-analytic model and the model of \citep{IP1982} when $\bar g < 10^2$ for the case of $M=10~\mr{M}_{\odot}$,
$\bar g < 1$ for the case of $M=10^{3}~\mr{M}_{\odot}$ and $\bar g < 10^{-2}$ for the case of $M=10^{5}~\mr{M}_{\odot}$. At larger accretion rates our model overproduces luminosity at sufficiently high
frequencies, probably, due to the neglect of synchrotron cooling of electrons. Combining these results we use below the condition that ${\bar g M\over 10^{3}~\mr{M}_{\odot}} < 1$  as a criterion of validity of our model, which results in the following condition 
for the mass accretion rate
\begin{equation} 
\dot m < 1600~\beta_V.
\label{e10}
\end{equation}  

Note, however, that for our purposes it is important to have reliable estimates of radiation flux at frequencies corresponding  to the minimal values of MM and JWST sensitivity curves, which are $2.8\cdot 10^{12}~\mr{Hz}$ and $8.6\cdot 10^{13}~\mr{Hz}$, for MM and JWST, respectively, see the next Section. As seen from Figs. \ref{Fig1} and \ref{Fig2}, at these frequencies the agreement between our model and that of \citep{IP1982} extends to somewhat larger values of mass flux, so in what follows
we are going to imply a less stringent constraint assuming that our model is valid when $\dot m \le 10^{4}$. 
 
That our very simple model is able to reproduce results of more complicated numerical calculations implies that the formation of the spectrum rather weakly depends on details of several effects implemented in numerical models, such as e.g. relativistic effects or the dependency of electron temperature on radius, provided that accretion rate is sufficiently small.

\subsection{Semi-analytic expressions for spectrum and luminosity for the case of small optical depth}
\label{SA}

\begin{figure}
\begin{center}
\includegraphics[width=14.0 cm]{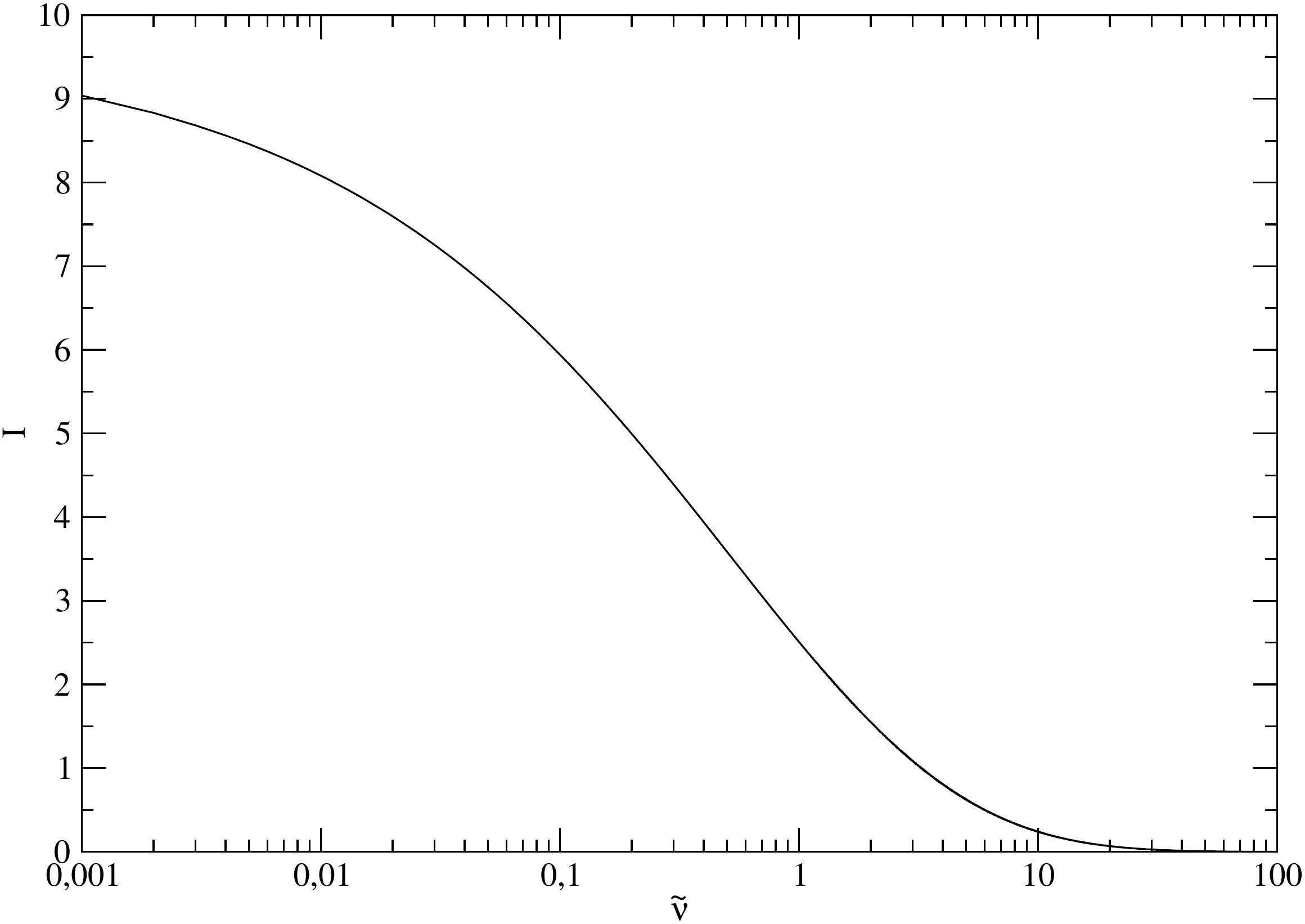}
\end{center}
\caption{The result of evaluation of the integral $I(\tilde \nu)$ as a function of the dimensionless frequency $\tilde \nu$.}
\label{FI}
\end{figure}

When the effects determined by synchrotron optical depth $\tau $ can be neglected, which is always valid at
sufficiently high frequencies, one can develop a semi-analytic
approach to the calculation of $L_{\nu}$. Moreover, since total luminosity $L_{tot}=\int d\nu L_{\nu}$ is 
mainly determined by these high frequencies, we can obtain an approximate analytic expression for it in 
this limit. For that we neglect the factor $e^{\tau(R^{'})}$ in the expression for $L_{\nu}$ in (\ref{e4}) and 
change the integration variable in this expression from $r$ to $x=\nu/\nu_{c}$, see equations (\ref{e5}) and
(\ref{e6}). We have
\begin{equation}
L_{\nu}={4\over 13}L_{0}{\tilde \nu}^{-1/13}I(\tilde \nu), \quad I(\tilde \nu)=\int^{\infty}_{({3\over 2})^{13/4}
\tilde \nu}dx x^{-12/13}\Phi(x),
\label{nt1}
\end{equation}   
where $L_{0}=L_*\beta_{B}^{1/2}(\beta_V \dot m m)^{3/2}$. The quantity $I(\tilde \nu)$ can be evaluated numerically, with Fig. \ref{FI} showing the result. 
One can see from this figure that $I(\tilde \nu)$ decreases monotonically with $\tilde \nu$, it is approximately equal 
to $9$ when $\tilde \nu =0$  which was checked numerically. Thus, when $\tilde \nu \ll 1$, but is still sufficiently high for our approximation of zero optical depth to be valid $L_{\nu}\approx {36\over 13}L_{0}{\tilde \nu}^{-1/13}$, see 
also e.g. \citep{BK}, \citep{IP1982} and references therein. The total luminosity is given by the expression
\begin{equation}
L_{tot}={4\over 13}L_{0}\nu_{0}\int^{\infty}_{0}d\tilde \nu {\tilde \nu}^{-1/13}I(\tilde \nu),
\label{nt2}
\end{equation}  
where we formally extend integration limits from $0$ to $\infty$ and $\nu_{0}=\nu_{c}(r=1)=\nu_*\epsilon^2 (\beta_B\beta_V^{-1}\dot m m^{-1})^{1/2}$. It can be checked that the integral in (\ref{nt2}) approximately equals to
$12.4$ and we finally obtain
\begin{equation}
L_{tot}=9.5\cdot 10^{28}\epsilon^{2}\beta_{B}\beta_{V}{\dot m}^{2}m ~\mr{erg~s^{-1}}=3.8\cdot 10^{27}\epsilon_{0.2}^{2}\beta_{B}\beta_{V}{\dot m}^{2}m~\mr{erg~s^{-1}},
\label{nt3}
\end{equation}  
where $\epsilon_{0.2}=\epsilon/0.2$ and we remind that we use $\epsilon=0.2$ in our calculations. Note 
that the expression for luminosity (\ref{nt3}) is different from that of \citep{IP1982}. This is
because \citep{IP1982} considered the regime of strong synchrotron cooling of electrons, which is realised
at accretion rate larger than those considered in this Paper. Therefore, our expression for $L_{tot}$ and
that of \citep{IP1982} complement each other. 

\section{The possibility of detection of IBHs with MM and JWST}
\label{detection}

In order to find under which conditions IBHs can be detected by MM and JWST we need to know a typical density of
the interstellar medium and a typical velocity of an IBH, which enter equation (\ref{BH}), a typical number of
IBHs, $N_{IBH}$, within a region of GC potentially interesting for observations as well as sensitivity curves of 
both instruments. The latter, in turn, are determined by internal characteristics of these instruments, 
background radiation of the region and, in case of MM, which has a worse spatial resolution than JWST, by 
the effect of confusion between radiating IBHs and distant IR galaxies. Note that equation (\ref{BH}) also 
contains the accretion efficiency parameter $\lambda$, which is treated below as free.

\subsection{Properties of interstellar medium and IBHs population relevant to the problem}

We remind that we consider both possibilities that IBHs are of primordial origin and were born as a result of stellar
evolution. In the former case it is reasonable to assume that their density is a fraction, $f$, of the density of 
dark matter, $\rho_{DM}$. The value of $\rho_{DM}$ at the distance of the order of $100$~pc from GC depends significantly on
whether its distribution has a core or a cusp in the centre of our Galaxy. If the former assumption is correct
$\rho^{core}_{DM}\approx 4\cdot 10^{-2}~\mr{M}_{\odot}~\mr{pc}^{-3}$ (see e.g. \cite{Nesti}), while the latter one yields $\rho^{cusp}_{DM}\approx 1.9 ~\mr{M}_{\odot}~\mr{pc}^{-3}$ (see e.g. \cite{McMillan}. Accordingly, we distinguish
between $f^{core}$ and $f^{cusp}$ noting that in the latter case our results could give  stringent constraints 
on PBHs abundance in case of non-detection. In the case of IBHs born as a result of stellar evolution their possible 
number densities depends on the value of kick velocity during the supernova explosion and their formation rate. \cite{Tsuna2018} recently investigated 
such a problem. From their results it follows
that for a typical BH mass equal to $10~\mr{M}_{\odot}$ the number density is $\sim 10^{-1}~\mr{pc}^{-3}$ { (D. Tsuna,
private communication.)} and, accordingly, 
we represent the number density of stellar evolution IBHs in the form $n_{IBH}=f^{SE}10^{-1}~\mr{pc}^{-3}$. Note that unlike $f^{core}$ and $f^{cusp}$,
$f^{SE}$ could be larger than one. 

As a model of interstellar medium we employ the model of 
distribution of molecular hydrogen $H_2$ within the Central Molecular Zone obtained in paper \cite{Fer}, who describe 
it as a flattened and elongated toroidal structure inclined with respect to the line of sight. The number density of hydrogen is estimated to peak at $150~\mr{cm}^{-3}$ and decays exponentially in vertical and radial direction with typical lengthscales $18$~pc and $137$~pc, respectively. We define the volume of CMZ, $V_{CMZ}$, as the boundary of the
region, where the number density is larger than $50~\mr{cm}^{-3}$ and use the analytical expression from \cite{Fer} to estimate it. We get $V_{CMZ}=2.7\cdot 10^{6}~\mr{pc}^{3}$. As a typical density inside this volume we take
$10^2~\mr{cm}^{-3}$, which results in the mass density $\rho=3.34\cdot 10^{-22}~\mr{g~cm}^{-3}$  in eq. (\ref{BH}). 

We assume that velocity distribution of IBHs is the same as that of the bulge stars at the appropriate distances from
GC. In \cite{Valenti2018} it is shown that the distribution of radial velocities can be approximately modelled by the Maxwellian one, with radial dispersion $\sim 130~\mr{km~s^{-1}}$. The dispersion of absolute values of velocity $v$,  $\sigma$, should be $\sqrt{3}$ times larger, so we use $\sigma=225~\mr{km~s^{-1}}$. Note that equation (\ref{BH}) also contains the sound speed $c_s$. Taking into account that a typical temperature of the $H_2$ component of CMZ is expected  to be small, of the order of $100$~K, the sound speed  { $v_{s}\approx 0.5~\mr{km~s^{-1}}$} is expected to be smaller than the typical velocities of interest and is neglected from now on.

\subsection{The sensitivity curves of MM and JWST}   

The sensitivity of JWST was extracted from its exposure time calculator { , see https://jwst.etc.stsci.edu/,} for NIRcam and MIRI imaging. The sensitivity of MM was computed assuming that it will have a 10~m mirror with emissivity of~0.05, surface RMS accuracy of 10~microns and temperature 20~K. The sensitivity depends on the level of detector noise, the emission of telescope construction and the sky background. The detectors are characterised by the Noise Equivalent Power NEP$=10^{-19}$~W~Hz$^{-1/2}$. The background sky emission is calculated using the dust model from \cite{planck_dust}. The dust parameters were averaged over a circle of radius 200~pc around the Galactic centre. The spectral resolution is set to $\lambda / \Delta \lambda = 5$.

The fact that our Galaxy is transparent in the far infrared band, with the optical depth below 0.01 even close to its centre, on one hand enhances our ability to search for IBHs in this band, but on the other hand, leads to a serious limitation. The high density of distant submillimetre galaxies on the sky together with relatively low resolution of even a 10-m submillimetre telescope result in the effect of confusion: below a certain level of flux density images of point sources on the sky start to overlap. This critical level of flux density depends on the telescope diameter and on wavelength. For MM it was estimated in \cite{ermash18}, based on N-body cosmological simulations and a model of galactic emission, and we use this estimate in the present work. For wavelengths shorter than 50 microns the confusion by distant galaxies disappears, so it is not important for JWST.

Also, the observations of the GC at wavelengths shorter than
30 microns are affected by extinction. We adopt the infrared extinction
towards the GC from \cite{Fritz11} and take it into account for computing
the JWST sensitivity.

One should note that even if we look for a source with a flux density below the confusion limit, it still can be observed if some criterion is used allowing to distinguish it from distant galaxies. Indeed, differences in the spectral energy distribution (IBHs have a power-law SED in submm range) and variability can be used to distinguish IBH candidates from extragalactic sources.
Distant galaxies, even if they host an accreting supermassive black hole, should have a plenty of spectral lines in the submillimeter range. These lines can be used to determine the redshift of a galaxy and other parameters, including AGN activity. They also can be used to get rid of the confusion, see
e.g. \citep{Raymond10}. The sensitivity of Millimetron in its spectral mode of observations
is sufficient to detect spectral lines of a galaxy with a flux density 2 orders of magnitude below the confusion limit at 300 microns.
The small mass of IBHs in comparison with supermassive black holes should result in a rapid variability of the observed flux density. Actively accreting stellar-mass BHs show detectable variations on very large range of frequencies up to 0.1~Hz in X-rays \citep{Atapin} and 0.5~Hz in optical band \citep{Burenin}. In the submillimeter range, the SS433 microquasar core shows flux density variations of 20\% on the timescale of 10 minutes. Of course, these variation timescales are determined by processes
in an accretion disc, but a spherical accretion onto a black hole of a similar mass should have a comparable variability timescale. This feature could also make it possible to distinguish IBHs from much more massive and, therefore, much more slowly changing AGNs.

 Thus, we give all our estimates of IBH detection probability using two variants of MM sensitivity curves: limited by the detector noise and  the level of background in this region  and limited by the confusion effect.

The detector-limited sensitivity of both JWST and MM depends on time as $t^{-1/2}$. We assume that the CMZ will be studied in detail with both these observatories. A typical period of observational time for the key scientific programs of these observatories is about 500---1000~hours. The field of view of JWST NIRcam is 9.7~arcmin$^2$, for MIRI 2.3~arcmin$^2$, for MM the field of view at highest frequencies will be about 36~arcmin$^2$. This means that for the imaging of CMZ, a rough estimate of 1~hour integration time seems reasonable within an accuracy of an order of magnitude.

\subsection{Results}

\begin{figure}[h]
\begin{center}
\includegraphics[scale=0.5]{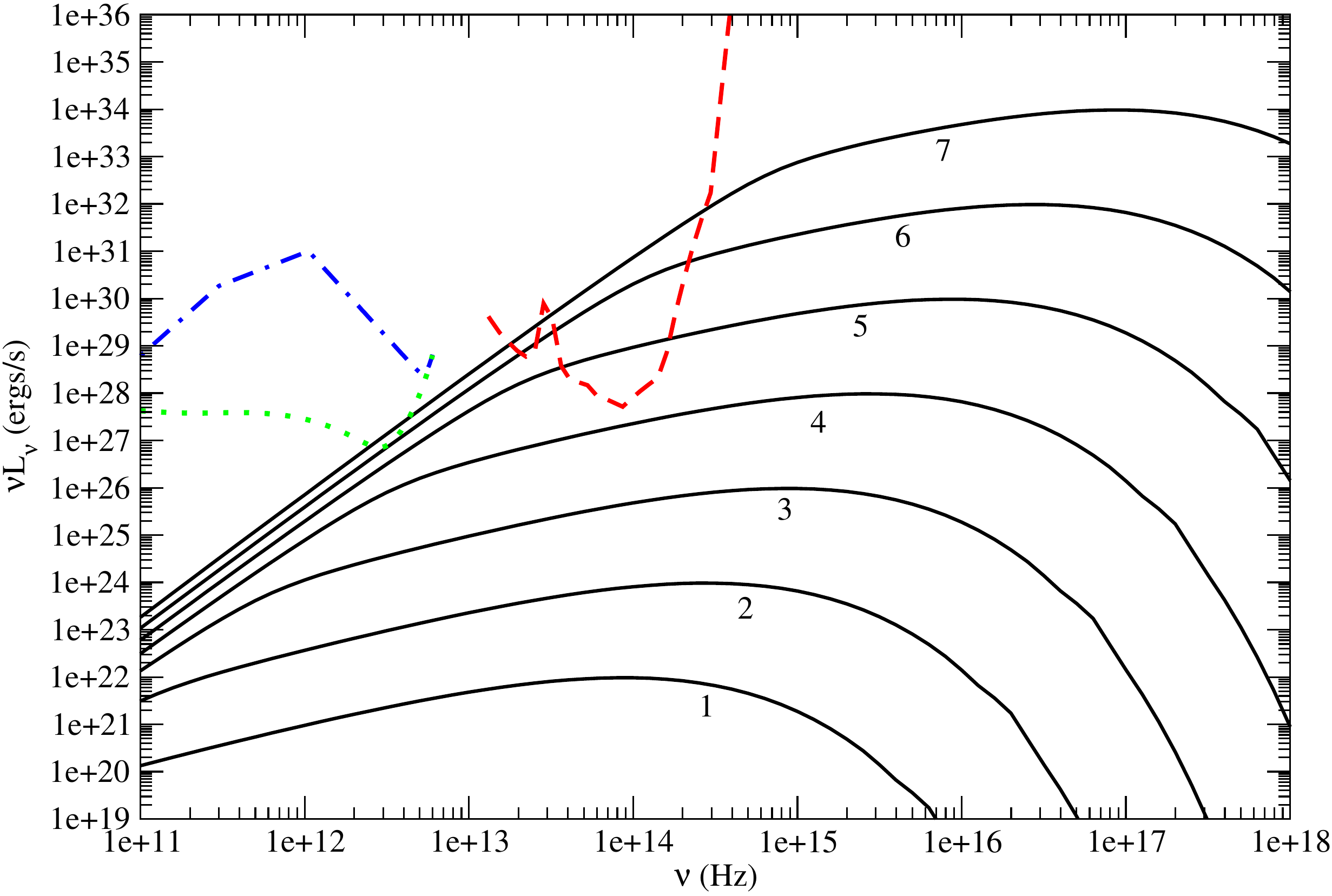}
\end{center}
\caption{Spectra of an IBH with $M=1~\mr{M}_{\odot}$ together with sensitivity curves of MM and JWST. The luminosity 
curves shown as solid curves, the dashed, dotted and dot dashed curves correspond to JWST sensitivity limit, MM sensitivity limit with the confusion effect being neglected and that of MM with the confusion effect being taken into account, respectively. The solid curves
with larger values { of their argument correspond to larger $\dot{m}$ , the labels 1,2,3,4,5,6 and 7 denote the curves with  $\dot{m}=10^{-2}$, $10^{-1}$,
$1$, $10$, $10^2$, $10^3$ and $10^4$, respectively.}}
\label{Fig4}
\end{figure}

\begin{figure}[h]
\begin{center}
\includegraphics[scale=0.5]{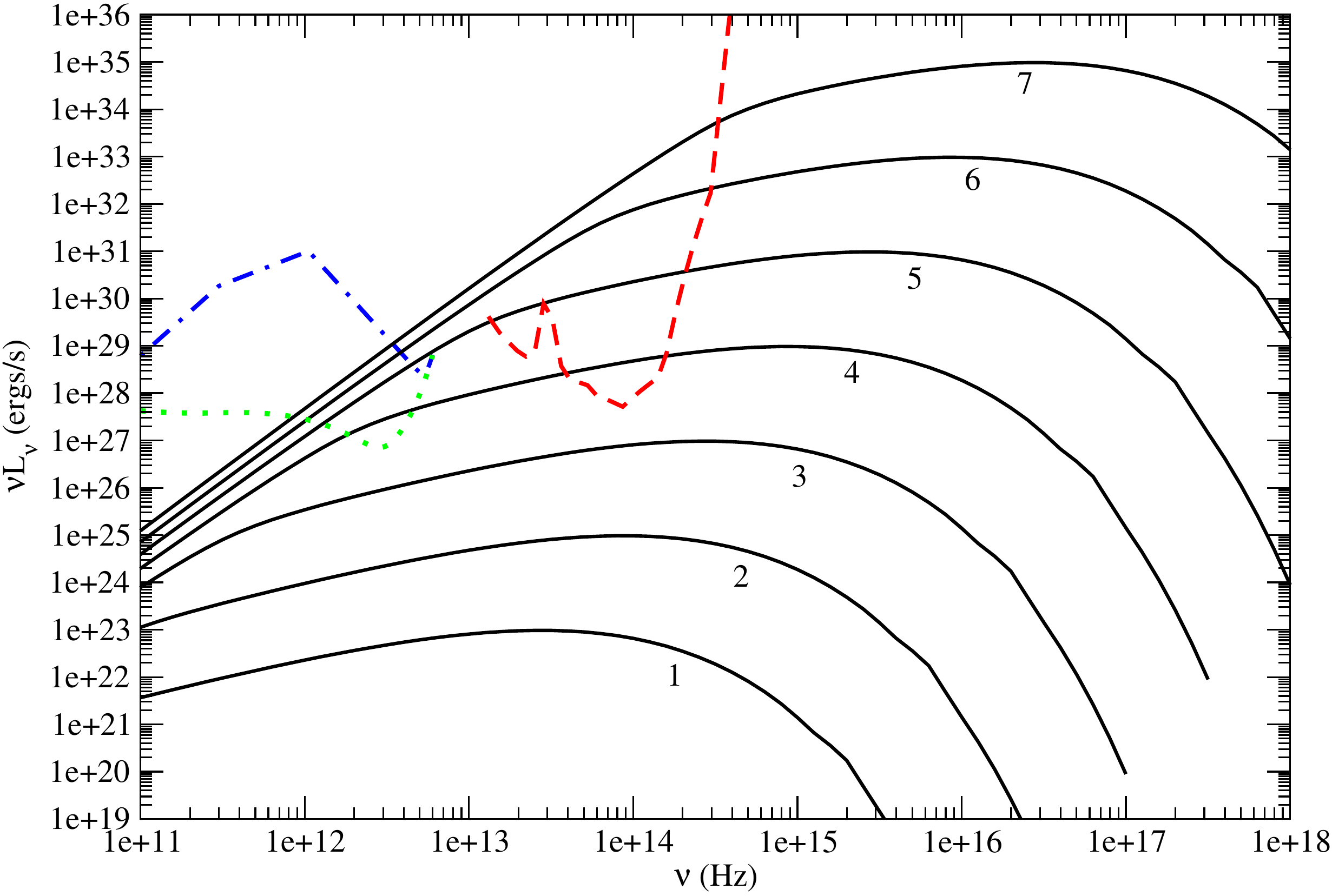}
\end{center}
\caption{Same as \ref{Fig4}, but for $M=10~\mr{M}_{\odot}$.}
\label{Fig5}
\end{figure}

\begin{figure}
\begin{center}
\includegraphics[scale=0.5]{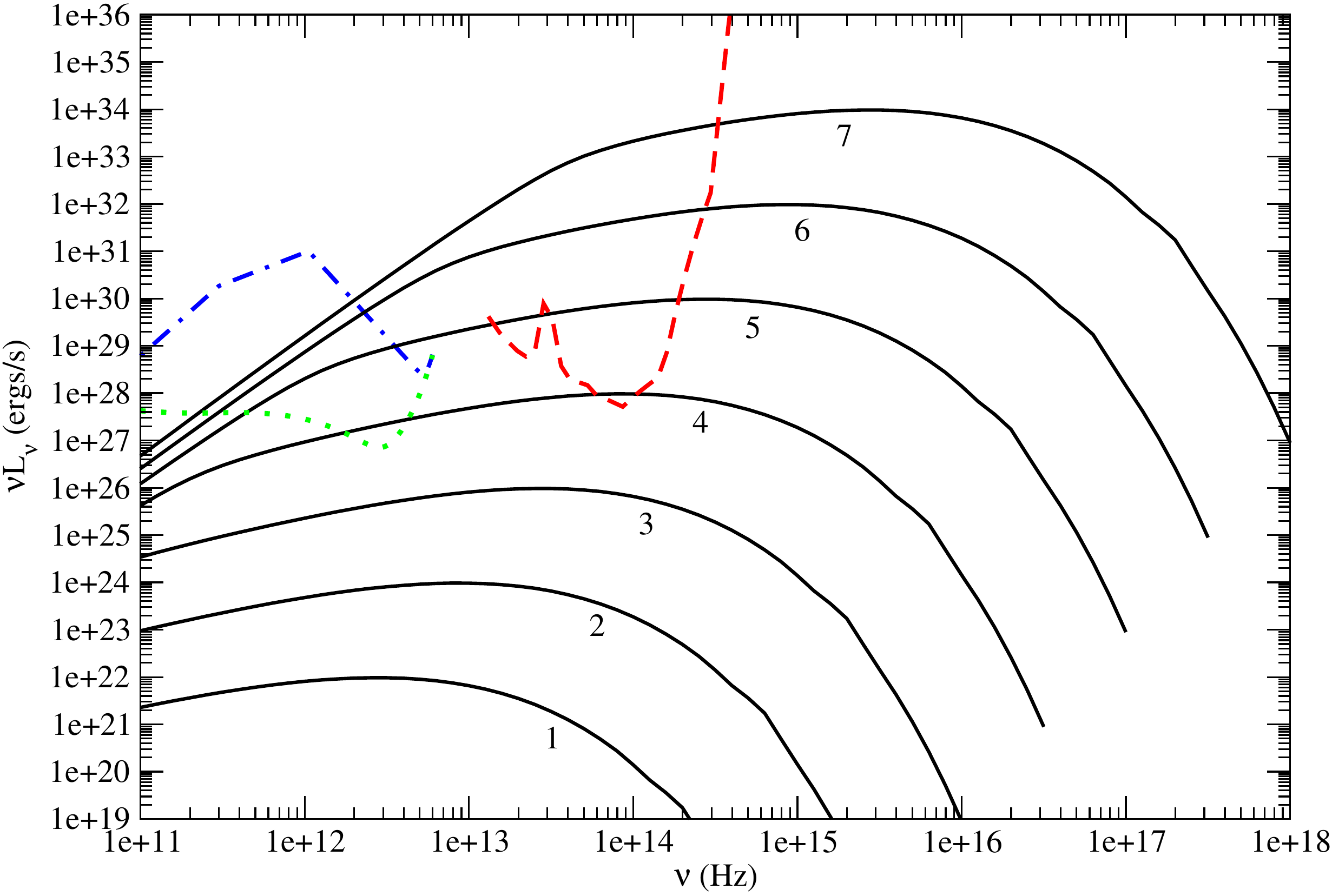}
\end{center}
\caption{Same as \ref{Fig4}, but for $M=100~\mr{M}_{\odot}$. { Now we show $\dot{m}$ in the range $10^{-3}$ − $10^3$, so the labels  1,2,3,4,5,6 and 7 denote the curves with  $\dot{m}=10^{-3}$, $10^{-2}$,
$1$, $10$, $10^2$, $10^3$ and $10^4$.}}
\label{Fig6}
\end{figure}

We show the dependencies of luminosities, $\nu L_{\nu}$, calculated in our model as functions of frequency 
in Figs. \ref{Fig4}-\ref{Fig6} together with MM and JWST luminosity curves. Figs. \ref{Fig4}-\ref{Fig6}
correspond to $m=0.1$, $1$ and $10$ ($M=1~\mr{M}_{\odot}$, $10~\mr{M}_{\odot}$ and $100~\mr{M}_{\odot}$), respectively,
with the luminosity curves shown as solid curves, the dashed, dotted and dot dashed curves corresponding to 
JWST sensitivity limit, MM sensitivity limit with the confusion effect being neglected and that of MM
with the confusion effect being taken into account, respectively. 
The solid curves with larger values correspond to larger
$\dot m$, every subsequent curve has $\dot m$ being multiplied by $10$ and we show $\dot m$ is the range 
$10^{-2}-10^{4}$ in case of $m=0.1$ and $1$, see Figs. \ref{Fig4} and \ref{Fig5}, and in the range $10^{-3}-10^{3}$
in case of $m=10$, see Fig. \ref{Fig6}. 

\begin{figure}
\begin{center}
\includegraphics[scale=0.5]{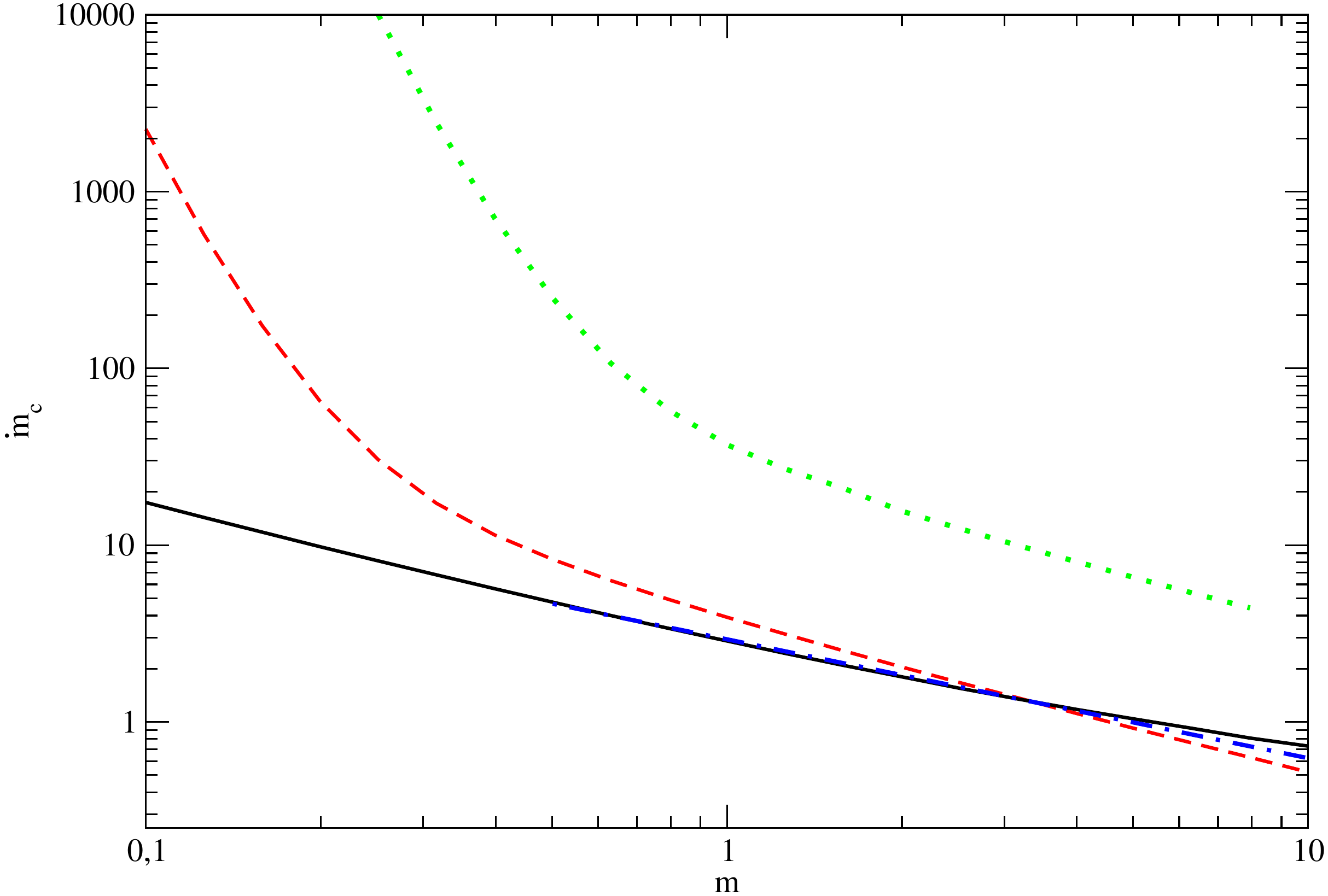}
\end{center}
\caption{The dependency of dimensionless critical accretion rate $\dot m_{c}$ on IBH's dimensionless mass $m=M/(10~M_{\odot})$. Solid, dashed and dotted curves correspond to JWST detection limit, that of MM without the confusion effect being taken into account, and that of MM with this effect being accounted, respectively.}
\label{Fig7}
\end{figure}

One can see from these figures that both instruments can detect IBHs provided that their masses and/or accretion rates are sufficiently large. JWST has a larger range of available $m$ and $\dot m$,  whereas in the case of MM the confusion effect rather significantly hampers observations and the corresponding range of $m$ and $\dot m$ is rather small. We quantify detection capabilities of JWST and MM introducing a critical accretion rate, $\dot m_{c}$, above which observations are possible, for a given value of $m$. The dependencies of $\dot m_{c}$ on 
$m$ are shown in Fig. \ref{Fig7}, for JWST sensitivity curve and MM sensitivity curves with and without 
confusion effect as solid, dashed and dotted curves, respectively. This figure demonstrates that in the case of JWST $\dot m$ depends on $m$ approximately 
as a power of $m$. We find that the JWST curve at  $0.5 \le  m \le 10$ agrees  with the expression
\begin{equation}
\dot m_{c}=Am^{-b}, \quad A=2.9, \quad b=0.67
\label{fit}
\end{equation}     
 within an order of or less than 10 per cent. This dependency is shown 
as the dot dashed curve in Fig. \ref{Fig7}\footnote{It is interesting to point out that a similar expression
can be obtained from our approach developed in Section \ref{SA}. Indeed, as seen from Figs. \ref{Fig4}-\ref{Fig6} a minimal value of JWST sensitivity curve is approximately $5\cdot 10^{27}~\mr{ergs~s^{-1}}$, which is
close to the numerical value of $L_{tot}$ given by eq. (\ref{nt3}) when $\dot m =m=1$. Neglecting factors 
order of unity and assuming that the total luminosity is mainly determined by frequencies close to that 
corresponding to the minimum of JWST sensitivity curve, we obtain from (\ref{nt3}) 
an estimate $\dot m_{c} \sim m^{-0.5}$, which is quite close to (\ref{fit}).}.    

\begin{figure}
\begin{center}
\includegraphics[scale=0.5]{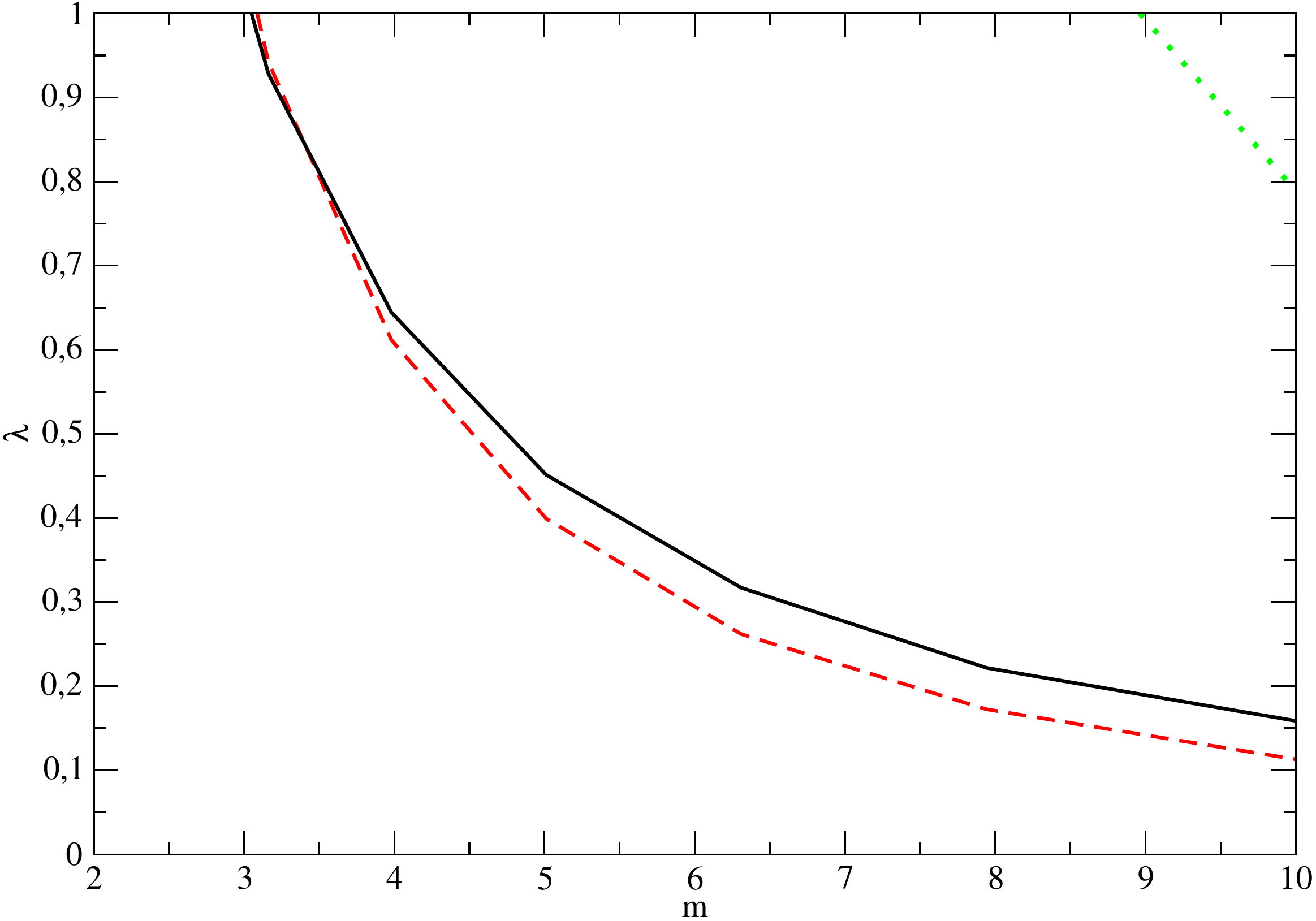}
\end{center}
\caption{The dependency of the parameter $\lambda $ on  $m=M/(10~M_{\odot})$. Curves of different styles describe the same cases as in 
Fig. \ref{Fig7}.}
\label{Fig8}
\end{figure}

\begin{figure}
\begin{center}
\includegraphics[scale=0.5]{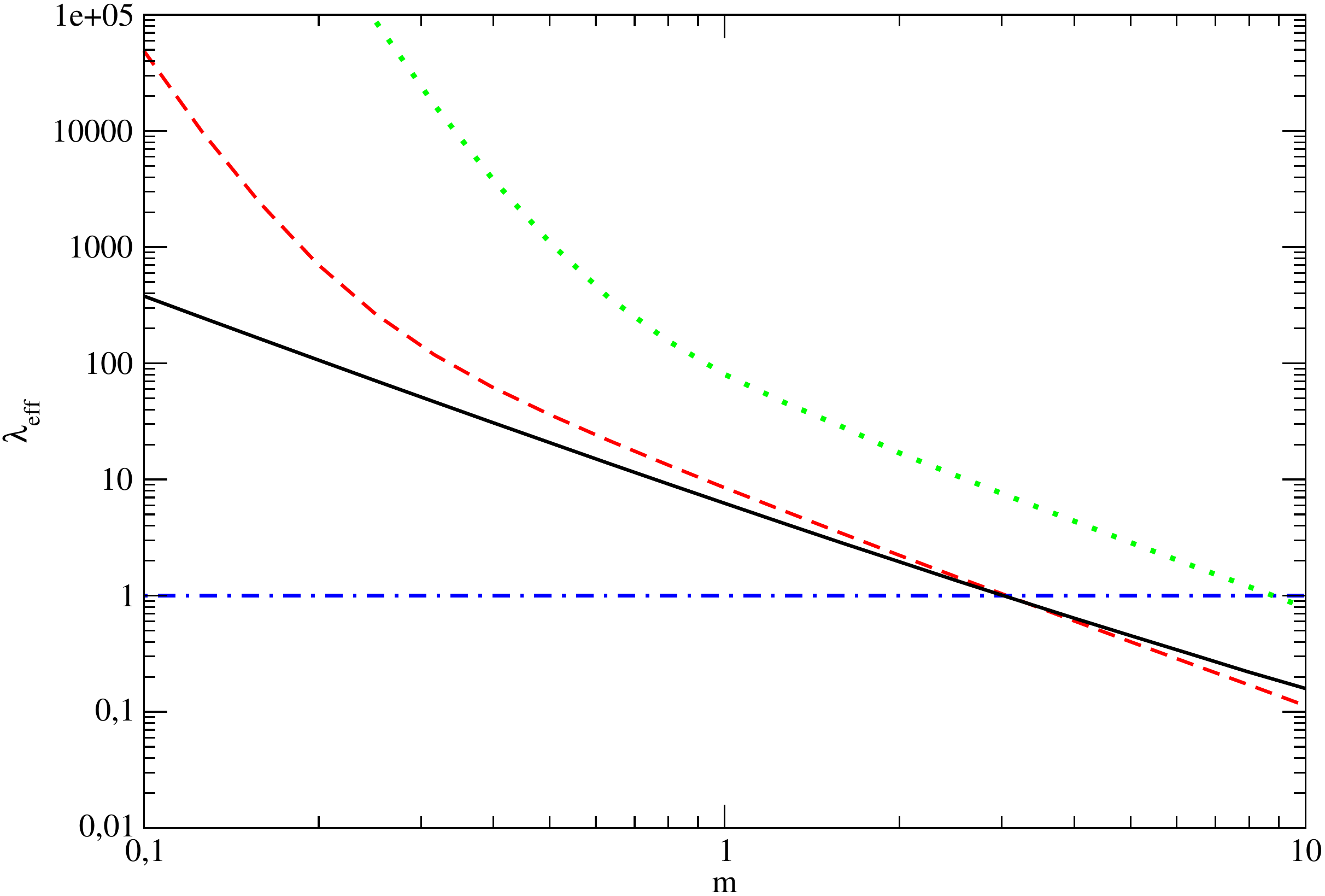}
\end{center}
\caption{Same as Fig. \ref{Fig8}, but now the dependency of parameter $\lambda_{eff} $ is shown.}
\label{Fig9}
\end{figure}  

\begin{figure}
\begin{center}
\includegraphics[scale=0.5]{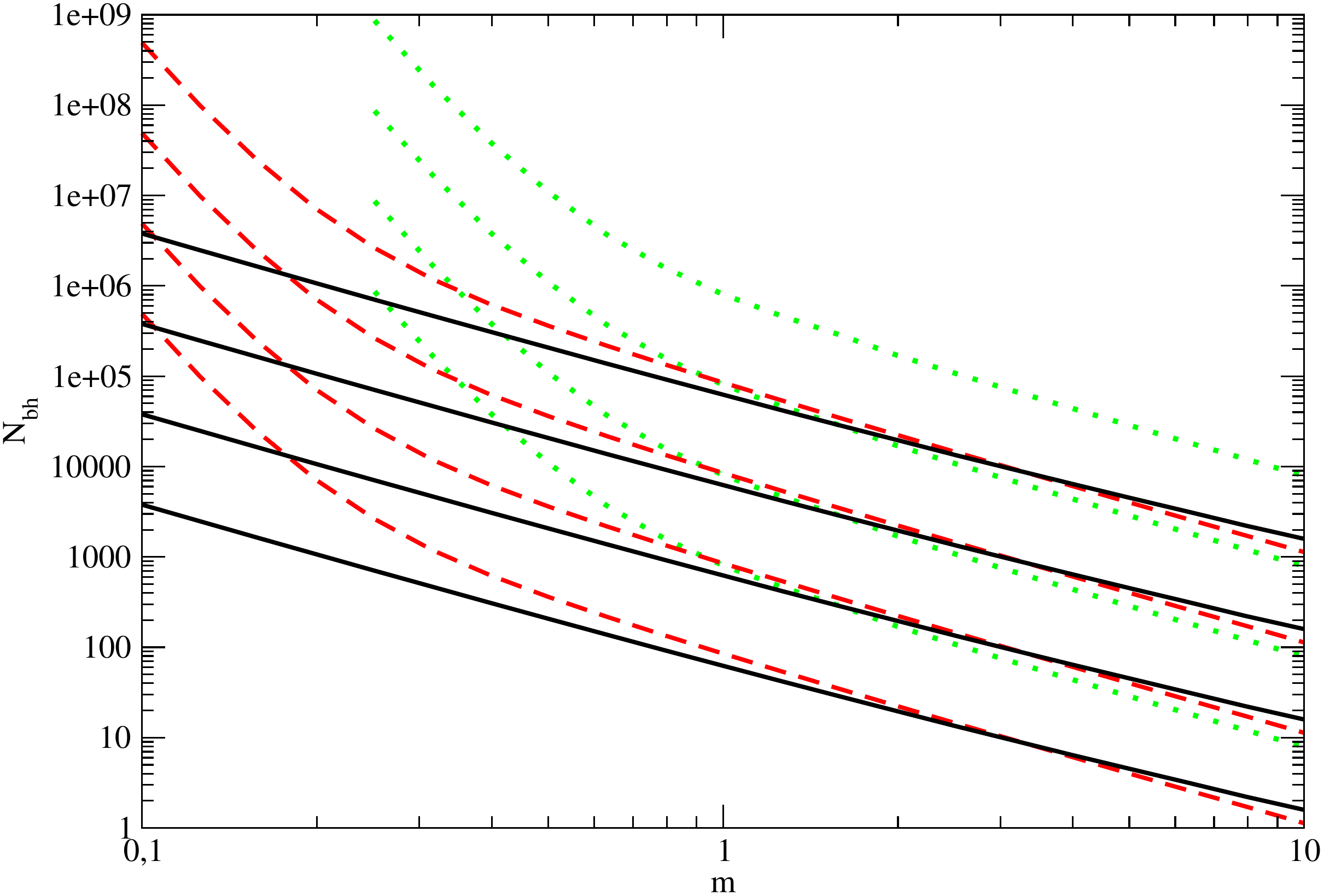}
\end{center}
\caption{The minimal number of black holes within CMZ needed for observations
with JWST and MM, for fixed values of the efficiency parameter $\lambda$, as a function of mass. Solid, dashed and dotted curves
corresponding to observations with JWST and MM without and with the confusion limit being taken into account. Curves with progressively larger values of $N_{bh}$ for a given value of $m$ correspond to progressively smaller values of $\lambda$.}
\label{Fig10}
\end{figure}

The accretion rate given by expression (\ref{BH}) should be smaller that its critical value for 
$\lambda \le 1$. This requirement sets a limit of the possibility of observations of IBHs by a given instrument
and for a given value of the density of the interstellar medium, IBH mass and velocity. In order to compare (\ref{BH}) 
with $\dot m_{c}$, we express the accretion rate given by (\ref{BH}) in terms of our characteristic 
unit: $\dot m=\dot M/(10^{-8}\dot M_{E})$, set  $c_s=0$ in (\ref{BH}), substitute there our chosen value of
density of interstellar medium $\rho=3.34\cdot 10^{-22}~\mr{g~cm}^{-3}$ and assume that a 'typical' IBH velocity
$\sigma=225~\mr{km~s^{-1}}$. We obtain
\begin{equation} 
\dot m=0.46 \lambda m{\tilde v}^{-3},
\label{e11}
\end{equation}  
where $\tilde v=v/\sigma$. 

To begin with let us consider a black hole travelling at the typical velocity and set $\tilde v=1$ in eq. (\ref{e11}).
In this case from the requirement that $\dot m > \dot m_{c}$ for $\lambda < 1$ we get lower limits on
black hole masses, $M_{crit}$, available for observations. For larger black hole masses the condition   $\dot m = \dot m_{c}$ gives a lower limit on the efficiency parameter $\lambda $, which is shown in Fig. \ref{Fig8}.
One can see from this figure that JWST and MM without taking into account the confusion effect could be able to observe IBHs with masses larger than $M_{crit}=30~\mr{M}_{\odot}$, while allowing for the confusion effect  MM is
able to detect IBHs only when $M > 90~\mr{M}_{\odot}$, respectively. The range
of available values of $\lambda $ is rather limited to $\lambda > \sim 0.1$ even for the largest considered 
black hole mass $100~\mr{M}_{\odot}$ when the confusion effect is neglected and $\lambda >  0.8$ in the opposite case. That $\lambda $ should be relatively large may be a problem for observations since some works suggest significantly
smaller values, see e.g. \cite{Igumen}. On the other hand, assuming that high accretion efficiency is sustained
and taking into account that the available volume of CMZ $\sim 10^{6}~\mr{pc}^3$, it could be possible to observe
IBHs with number density as small as $10^{-6}~\mr{pc}^{-3}$.

Available values of $\lambda $ and $M_{crit}$ can be smaller in a situation when there are many IBHs within
CMZ, with their number $N_{bh} \gg 1$ and, accordingly, their number density $\gg 10^{-6}~\mr{pc}^{-3}$. In this case assuming the Maxwellian statistics of distribution of IBHs over velocities we can estimate the number of IBHs having their velocity smaller than $\tilde v$, $N(\tilde v)$, as $N(\tilde v) ={\tilde v}^3N_{bh}$\footnote{Note that we assume that IBH velocities are defined with respect to GC and do not take into account that molecular clouds comprising CMZ have themselves  appreciable velocities with respect to that. However, a simple estimate shows
that our expression for $\tilde v$ remains valid by order of magnitude 
for the expected values of this velocity, which are smaller than or
order of $200~\mr{km~s^{-1}}$, see e.g. \cite{Sof}, \cite{Hen}.}. We need 
at least one IBH for observations, and, from the condition $N(\tilde v)=1$ we estimate its velocity as 
$\tilde v \sim N_{bh}^{-1/3}$. Substituting this value to (\ref{e11}) we obtain
\begin{equation} 
\dot m=0.46 \lambda_{eff} m, \quad \lambda_{eff}=N_{bh}\lambda.
\label{e12}
\end{equation}      
The condition of observability $\dot m \ge \dot m_{c}$ results in the lower limit on $\lambda_{eff}$, which is shown in Fig. \ref{Fig9}. Note that it is technically the same curve as shown
in Fig. \ref{Fig8}, but now the condition $\lambda_{eff} < 1$ is relaxed and we show the whole considered
range of $m$, depicting by the additional dot dashed horizontal line $\lambda_{eff}=1$ the previous range of values\footnote{Note that we have the obvious restriction $\lambda_{eff} \le N_{bh}$.}.  
Now, assuming that $\lambda \sim 1$ 
we find that JWST would be able to detect an IBH even with $M\sim 1\mr{M}_{\odot}$ provided that $N_{bh} > 4\cdot 10^2$ while observations with MM in this mass range require significantly larger $N_{bh}\sim 10^{4}-10^{5}$. 

Let us now consider smaller values of $\lambda$ setting, for definiteness, $\lambda=10^{-1}$, $10^{-2}$, $10^{-3}$
and $10^{-4}$. Using equation (\ref{e12}) and the condition  $\dot m \ge \dot m_{c}$ we obtain the 
lower limits on the number of $N_{bh}$ within CMZ as functions of $m$, for a given value of $\lambda $ and
each observational instrument. We plot the results in Fig. 
\ref{Fig10} with solid, dashed and dotted curves corresponding to observations with JWST and MM without and with the confusion limit being taken into account. Curves with progressively larger values of $N_{bh}$ for a given value of $m$ correspond to progressively smaller values of $\lambda$.   

Expression (\ref{fit}) allows us to obtain an analytic expression for
$N_{bh}$ and, accordingly, a lower bound on black holes number density
$n_{bh}$ and mass density $\rho_{bh}$ for observations with JWST and $m \ge 0.5$.
Using (\ref{e12}) we have
\begin{equation} 
N_{bh}\approx 6.3\cdot 10^3\lambda^{-1}_{-3}m^{-1.67}, \quad n_{bh}\approx 2.4\cdot 10^{-3}\lambda^{-1}_{-3}m^{-1.67}~\mr{pc}^{-3}, \quad \rho_{bh}\approx 2.4\cdot 10^{-2}\lambda^{-1}_{-3}m^{-0.67}\mr{M}_{\odot}~\mr{pc}^{-3},
\label{e13}
\end{equation}  
where $\lambda_{-3}=\lambda/10^{-3}$ and we remind that we have adopted the 
volume of CMZ to be equal to $2.7\cdot 10^{6}~\mr{pc}^3$. Recalling our definitions
for the fractions of IBHs with respect to the mass density of dark matter 
in case of core and cusp, $f^{core}$ and $f^{cusp}$, respectively, and also the definition of the fraction of IBHs number density to the one expected as a result of stellar evolution, $f^{SE}$, we obtain from eq. (\ref{e13})
\begin{equation} 
f^{core}\approx 5.8\cdot 10^{-1}\lambda^{-1}_{-3}m^{-0.67}, \quad f^{cusp}\approx 1.2\cdot 10^{-2}\lambda^{-1}_{-3}m^{-0.67}, \quad f^{SE}\approx 2.4\cdot 10^{-2}\lambda^{-1}_{-3}m^{-1.67}.
\label{e14}
\end{equation}
Thus, even when the efficiency of accretion onto a PBH is rather small, 
$\lambda \sim 10^{-2}$ and there is a cusp  in the distribution of dark matter JWST will be able to detect PBHs if they were responsible for a substantial fraction of LIGO events. Its sensitivity is also sufficient to detect IBHs with the number density expected from the stellar evolution according to modern simulations. The limits get better when $m$ increases and improve dramatically for an efficient mode of accretion with $\lambda \sim 1$.

\section{Discussion}
\label{discussion}

\subsection{Summary and conclusions}

In this Paper we study the possibility of detection of isolated black holes
(IBHs) with stellar masses $M \sim 1-100~\mr{M}_{\odot}$  travelling within the so-called Central Molecular Zone (CMZ) at distances $\sim 100-200$~pc in the submillimetre and infrared range with the help of two planned space observatories, Millimetron (MM) and James Webb Space Telescope (JWST). We develop a simple model of synchrotron emission of thermal electrons by an
accreting black hole in the spherical mode of accretion, which, in general,
requires several straightforward numerical integrations and is semi-analytic   
when the optical thickness with respect to synchrotron self-absorption is
negligible. It is checked that this model agrees quantitatively with a more complicated numerical model of \cite{IP1982} for the values of accretion rates and black hole masses of interest. We assume the equipartition condition 
between magnetic field energy density and  potential energy density. Also
it is assumed that the accretion rate onto the black hole is a fraction, $\lambda$, of the Bondi-Hoyle-Lyttleton value determined, in general, by
the density and sound speed of CMZ gas and velocity of the black hole with respect
to it. 

We use MM and JWST detector sensitivity curves adding to them the contribution of background emission of CMZ gas and, in 
the case of MM, consider the 
effect of confusion between the objects of interest and distant submillimetre 
galaxies. In case of JWST this effect may be shown to be unimportant.

When $\lambda \sim 1$ both JWST and MM can detect a single accreting black
hole within CMZ travelling at typical velocity $\sim 200~\mr{km~s^{-1}}$ provided that
its mass is larger than $30~\mr{M}_{\odot}$ and, in case of MM the confusion effect
is neglected. When this effect is taken into account MM would be able to 
observe only rather massive black holes with masses $\sim > 90~\mr{M}_{\odot}$.
Bearing in mind that observations of CMZ must be a part of observational 
programme of both instruments we conclude that in the case of efficient 
accretion JWST and MM (in case when the confusion effect is eliminated 
by some additional means, say, using different variability timescale and
different spectral energy distribution of IBHs and submillimetre galaxies)
would be able to observe a single accreting black hole within CMZ. Thus,
{ when $\lambda \sim 1$ it would be possible to observe a single sufficiently massive IBH within CMZ, and, therefore, }
a possible number density of IBHs within $\sim 100-200$~pc from Galactic 
Centre could be as small as the inverse volume of CMZ $V_{CMZ}\sim 10^{-6}~\mr{pc}^{-3}$ for their detection. This exceeds the corresponding number density required
to explain the LIGO event by mergers of IBHs of primordial origin.

If a number of IBHs within CMZ is large, observations are still possible even
when $\lambda $ is rather small. This is because velocity distribution of 
IBHs with respect to CMZ gas is likely to be Maxwellian and there could 
be IBHs with a small velocity even when the bulk of them has typical velocity of the
order of its dispersion value. Since the accretion rate sharply increases with decrease of velocity, there could be an IBH with luminosity exceeding 
the sensitivity limit.

{ JWST seems to provide a better opportunity for detection of IBHs than MM.
This is determined by two circumstances. Firstly, since JWST 
has higher observational frequencies, it has a better resolution and is,
therefore, free from the confusion effect. Secondly,  it is fortunate to have 
the best sensitivity at frequencies order of $10^{14}Hz$, which are close
to typical frequencies where an IBH has its maximal luminosity. Therefore, we
consider this option for JWST in detail.} Our finding are expressed 
in terms of fractions $f$ of some reference mass/number density of IBHs. We consider the case
when IBHs are assumed to be primordial and constitute a fraction of dark
matter mass density in the case of its distribution in the form of a core or a cusp denoting the corresponding fractions as $f^{core}$ and $f^{cusp}$, respectively.
We also consider a possibility when IBHs are formed as a result of stellar
evolution introducing a fraction , $f^{SE}$, of their number density in terms of a value expected
in a recent model of their distribution in the Galaxy. The corresponding results are summarised in equation (\ref{e13}). From this equation it follows
that when $\lambda=10^{-2}$ and $M=10~\mr{M}_{\odot}$ we have $f^{cusp}\sim 10^{-3}$. Thus, observations with JWST could detect IBHs constituting a fraction of dark matter sufficient to explain the LIGO events  when $\lambda$ is larger than $10^{-2}$ and dark matter distribution has a cusp.

\subsection{Efficiency of accretion and limits of applicability of our accretion model}

The value of $\lambda $ is very difficult to estimate because there are many
factors, which could influence it. Among them are a degree of turbulence
and a typical value of magnetic field of CMZ gas, possible velocity shears
and density gradients inside CMZ, etc. Since such estimates must necessarily
be done by numerical means the issues of computational resolution, potentially large Mach numbers and the necessity to survey a large parameter space of
the problem also come into play\footnote{Note that in case of very large Mach numbers an analytic approach
is also possible, see \cite{Bis1}.} . Although at the present time a comprehensive approach to this problem is absent it could be possible to make some preliminary considerations based on numerical studies of some limiting cases.  When the medium is assumed to be uniform and unmagnetised numerical  simulations suggest that $\lambda$ can be order of unity, see \cite{accretion_high}. This conclusion is presumably valid for black holes
travelling at velocities order of $\sigma=225~\mr{km~s^{-1}}$ much larger than other 
characteristic values of velocity. These are the
sound speed $v_{s}\approx 0.5~\mr{km~s^{-1}}$, the Alfven speed $v_{A}\sim 1~\mr{km~s^{-1}}$ calculated for the expected typical value of CMZ magnetic field order of $\sim 10^{-5}$~G (e.g. \cite{Yusef}) and a typical velocity of turbulent motion $v_{turb}\sim 10~\mr{km~s^{-1}}$, see e.g. \cite{Ginsburg}, \cite{Shetty}. On the other hand, the accretion radius $R_{acc}={2GM\over v^{2}}\approx 5\cdot 10^{12}~\mr{cm}$ for $m=1$ and $v=\sigma$ is smaller than a typical size of density inhomogenities and large scale velocity shear, which is assumed to be equal to $R_{inh}\sim 1pc$ from now on, see e.g. \cite{Shetty}.   
    
IBH velocity gets, however, smaller when we consider the tail of Maxwellian
distribution over velocities.  Remembering that in this case we estimate $v \sim N_{bh}\sigma$, where $N_{bh}$ is the total number of IBHs in CMZ and
using eq. (\ref{e13}) we have $v\approx 16\lambda_{-3}^{1/3}m^{0.53}~\mr{km~s^{-1}}$ 
and $R_{acc} \approx 2\cdot 10^{14}$ when $v=16~\mr{km~s^{-1}}$ and $m=1$. The obtained value of velocity is already of the order of $v_{turb}$. Unfortunately, we
are not aware of any numerical calculations of accretion onto a black hole 
moving through a turbulent medium with a velocity of the order of an average velocity 
of turbulent motion. However, a similar case of accretion onto a stationary 
black hole with the average value of the turbulent velocity order of the sound speed 
considered in \cite{Turb} suggests that in this case $\lambda $ could be much 
smaller that one, with typical values $\sim 10^{-2}-10^{-3}$, which, however,
is already taken into account in (\ref{e13}) and (\ref{e14}), where 
it is implied that $\lambda \sim 10^{-3}$. 

Even when the accretion radius is much smaller than a typical size of inhomogenities it was suggested that even a small density/velocity gradient
could significantly influence accretion and lead to formation of an accretion
disc, see \cite{Illarionov1}. Later this problem was considered numerically, 
see e.g. \cite{Riffert}, where it was shown that, in fact, the amount of angular
momentum accreted to small distances due to the presence of these inhomogenities
is smaller than the analytic estimate. Also, the accreted angular momentum changes sign with time. Either of these effects could inhibit formation of a disc. Also, note that even if a disc is formed, it must have the
properties of the so-called Radiatively Inefficient Accretion Flow, which has many similarities with the spherical mode
of accretion considered in this Paper and could be treated by similar means, see e.g. \cite{Nar}.

\subsection{The possibility of radiation emission in other parts of spectrum in our model}

{ When the typical accretion rate is moderate, $\dot m \sim 1$, in our model the bulk of energy is released in the infrared and visible parts of
the spectrum. However, from Figs. \ref{Fig4}-\ref{Fig6} it follows that when $\dot m$ is relatively large, a significant fraction 
of radiation is emitted in X-rays and can be, in principle, detected by the modern X-ray space observatories such as 
\textit{XMM-Newton} and \textit{Chandra}. Let us estimate the possibility of detection of our sources in X-rays. For that
we note that the adopted  values for the concentration $n_{H_2}=100~\mathrm{cm^{-3}}$ in CMZ and its characteristic  size $d\sim100$~pc allow us to estimate   the  optical depth for an X-ray photon: $\tau=\sigma_{H_2}N_{H_2}$, where $\sigma_{H_2}=4.5\times10^{-23}~\mathrm{cm^2}(E/1~\mathrm{keV})$, $N_{H_2}\equiv n_{H_2}d\approx3\times10^{23}~\mathrm{cm^{-2}}$, and $E$ is the photon energy.  The CMZ becomes optically thin at $E\sim1$~keV, and we use this value for further estimates. Current experiments such as \textit{XMM-Newton} and \textit{Chandra} can observe sources with fluxes at these energies  $\sim10^{-15}~\mathrm{erg~cm^{-2}~s^{-1}}$, which translates to the threshold value of the X-ray luminosity $L_{x-ray}=10^{31}~\mathrm{erg~s^{-1}}$. The corresponding accretion rate $\dot{m}_{x-ray}$ is fairly stable in the whole mass range 1-100 $M_{\odot}$, gradually increasing from 700 to 1100. Such high accretion rates reside almost at the boundary of the region of validity of our model (see Eq. (11)) and in any case they are much higher than the accretion rates needed for detection in the sub-mm range $\dot{m_c}$. That means that IBHs that can be detected both in (sub)mm and  X-ray bands comprise only a small subset of IBHs detected only in sub-mm. The ratio of these subsets is determined by the ratio  $\dot{m_c}/\dot{m}_{x-ray}$, and changes from $\frac{1}{70}$ for the lightest considered IBHs (1~$M_{\odot}$))  to $\frac{1}{2000}$ for the heaviest IBHs with $M=100~M_{\odot}$.}

{ It is also important to note that radiation in harder parts of the spectrum  could originate, in principle, from the processes 
not considered in the Paper, such as the Compton scattering between synchrotron photons and thermal electrons 
and magnetic reconnection events leading to formation of a tail of high energy non-thermal electrons.} However, the $y$-parameter defined as $y=\int^{\infty}_{R_{min}}dR \sigma_{Th}n{k_BT_e\over m_ec^2}$, where $\sigma_{Th}$ is the Thomson cross section (\cite{Illarionov}) is extremely small in our
case, order of $\sim 10^{-5}$ telling that the Compton effects may be negligible.  The possible contribution of
non-thermal electrons was considered in \cite{Beskin2005}, who assumed that a fraction order of $0.1$ of released potential energy of infalling gas is transferred to relativistic electrons with a power-law distribution over their energy. They obtained rather flat spectra extending from infrared to gamma with a maximum of luminosity at frequencies of the order of $10^{15}$~Hz, which agrees with our approach.

\subsection{A comparison of our approach to other spectral models, which can be used for the search of IBHs} 

 In other papers considering similar problems of the search of accreting IBHs  
either the model of \cite{IP1982} (see, e.g. \cite{Mc1985})
or some phenomenological spectral models (e.g. \cite{Fender}, \cite{Gag}) have mainly been used. The former papers 
have dealt, however, with different observation sites like nearby molecular clouds in case of \cite{Mc1985}. Additionally, 
they have not considered the observational capabilities of MM and JWST. The latter papers used phenomenological 
relations between accretion rate and luminosities in different wave bands established either for binary systems 
or Active Galactic Nuclei. It is not clear to what extent these relations are valid for IBHs.

In particular, our conclusion that, for typical accretion rates, the maximal luminosity is expected to 
correspond to the optical and infrared bands seemingly contradicts 
the results of the recent paper \cite{Gag} who assume that 30 per cent of radiation is released in X-rays and
use the fundamental plane relation to estimate luminosity in the radio band. Note, however, that the 
approach undertaken in this paper relies on the assumption that radiation
forms in an accretion disc with properties similar to those found in much more luminous systems observed so far.
In this connection we note that in the model of  \cite{Beskin2005} luminosity in the harder part of spectrum appears to be an order of magnitude smaller than the region close to the maximum, thus, reconfirming the necessity to pay attention to the optical/infrared band for a search of IBHs.  

Also note that synchrotron emission can also arise in the bow shock, created by the
supersonically moving BH. This effect  was investigated in \citep{Wang2014}
in the particular scenario of intermediate mass black holes freely floating
in the Milky Way disk. Fig. 3 of \citep{Wang2014} can be used to show that
this mechanism is sub-dominant in the 1-100 $M_{\odot}$ mass range and the
radiation flux from the bow-shock is 5-6 orders of magnitude lower than the
sensitivity threshold even in the most optimistic case.

\subsection{Other possible sites for observation of IBHs}

{ IBHs could also be sought for in other locations, such as e.g. Giant Molecular Clouds (GMCs).  Let us estimate the possibility
of detection taking as an example the Orion Molecular Complex, Ori A and Ori B, which has its typical density $n\sim 10^{2}~\mathrm{cm^{-3}}$, size $R\sim 10~$pc, mass $M_{H_2}=10^{4}~M_{\odot}$, and distance $d=500~$pc, see e.g. \cite{Pet}.
We take into account that the accretion rate scales as
$$\dot{m}\approx \left(\frac{M}{10~M_{\odot}}\right)\left(\frac{n}{10^2~\mathrm{cm^3}}\right)\left(\frac{v}{300~\mathrm{km/s}}\right)^3$$
and calculate the  dependencies of luminosities $\nu L_{\nu}$ on frequency for  three typical values of IBH mass using 
$\dot m=1$. We show
these dependencies together with the 
relevant  MM and JWST sensitivity curves in Fig. \ref{Fig:nearby}. 
\begin{figure}
\begin{center}
\includegraphics[scale=0.5]{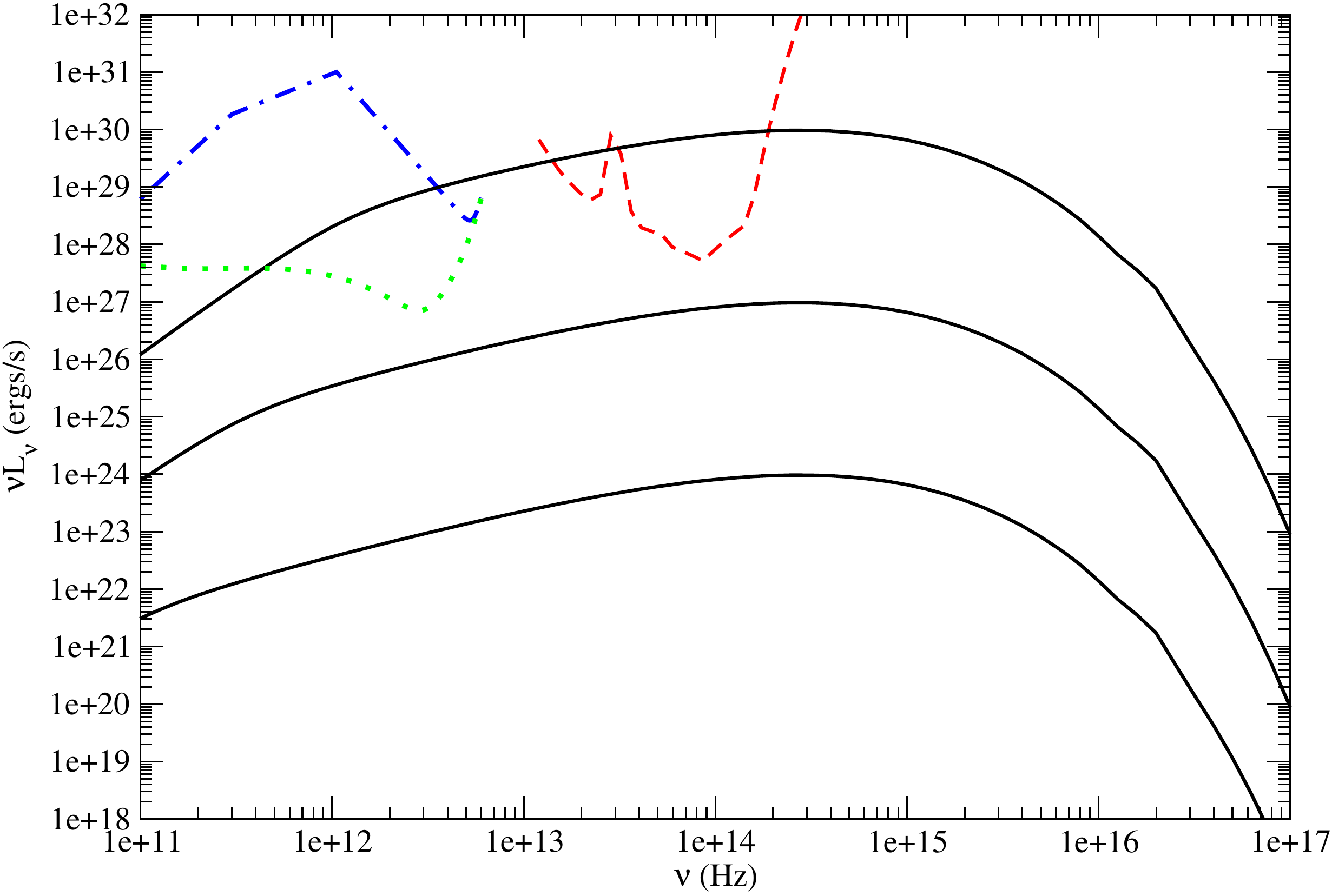}
\end{center}
\caption{Spectra of an IBHs with $M=1$, $10$ and $100~\mathrm{M}_{\odot}$ residing in Orion molecular cloud ($d=500~$pc),  together with  the relevant sensitivity curves of MM and JWST. The luminosity curves are shown as solid curves with larger values of
the argument corresponding to larger masses, the dashed, dotted and dot dashed curves correspond to the JWST sensitivity limit, the MM sensitivity limit with the confusion effect being neglected and that of MM with the confusion effect being taken into account, respectively.}
\label{Fig:nearby}
\end{figure}
The total number of black holes residing inside the MC can be estimated as
$N_{BH}=3f^{DM}\frac{10~M_{\odot}}{M_{BH}}$, so it is marginally possible to detect IBHs in the nearby GMC provided that their masses are not too small and they comprise a significant fraction of dark matter. In contrast to the CMZ case IBH candidates could be observed in optics due to much lower extinction in this regions. However, since the typical volume of a GMC $\sim 10^{3}~pc^3$ is much smaller than that of CMZ, the latter region appears to be preferable for the search of IBHs.}

\subsection{A possible development of our approach}
We assume the simplest possible model of CMZ. However, the region with distance order of $100-200$~pc from
the centre of our Galaxy is a very complex system. In particular, it contains extended molecular clouds
with much larger number densities, order of $10^{4}~\mr{cm}^{-3}$. It is of interest to repeat calculations reported
in this Paper for a realistic model of CMZ. Obviously, there some other sites to look for IBHs such as 
giant molecular clouds off GC or some other preferable places. Say, there are models of distribution
of compact remnants of stellar evolution in our Galaxy suggesting that they are distributed in a ring of
size $5$~kpc, see \cite{Pop1}, \cite{Pop}.

The CMZ is also populated with a prominent population of infrared sources: young stellar 
objects (YSO). At the distance of CMZ they will
look like point sources for both JWST and MM. YSOs possess a large variety of SEDs, some of them
may visually look quite similar to those of IBHs \citep{Robitaille06,Robitaille17}. So there is one more issue to be addressed: 
by future studies how to 
distinguish them. A more detailed comparison of spectra of YSOs and IBHs can help to answer this question.
Finally, the number of faint YSOs in the CMZ is not known. If the number of sources detectable by MM is 
of the order of $10^5$ or higher, they will create the confusion noise, similar to distant galaxies, and the sensitivity
of MM may be worse than expected in CMZ. A possible way to suppress the confusion considerably would be to use a difference in patterns of spatial distribution between YSOs and IBHs -- while the former reside overwhelmingly inside dense giant molecular clouds, the latter objects are distributed almost uniformly across the CMZ.

\section*{Acknowledgements}

We are grateful to E. V. Mikheeva, S. B. Popov and Ya. A. Shchekinov for important discussions and 
to V. N. Strokov and D. Tsuna for useful comments. This work was supported in part by RFBR grant 17-52-45053
and in part by RFBR grant 19-02-00199.  MSP acknowledges support of 
Leading  Science School MSU (Physics of Stars, Relativistic Compact Objects and Galaxies) and support  by the Foundation for the Advancement of Theoretical Physics and Mathematics ``BASIS'' grant 18-1-2-51-1.


\end{document}